\documentclass[journal]{IEEEtran}
\makeatletter
\def\BibTeX{{\rm B\kern-.05em{\sc i\kern-.025em b}\kern-.08em
		T\kern-.1667em\lower.7ex\hbox{E}\kern-.125emX}}
\makeatother
\usepackage{subfigure}
\usepackage{graphicx, amssymb, amsbsy, amsmath}
\usepackage{subfigure}
\usepackage{multirow}
\usepackage{cite,url}
\usepackage{color,soul}
\usepackage[font=footnotesize,labelfont=sf,textfont=sf]{caption}
\usepackage[lined,ruled,commentsnumbered]{algorithm2e}
\usepackage{algorithmic}
\DeclareGraphicsExtensions{.eps}

\usepackage{booktabs}
\def\squareforqed{\hbox{\rlap{$\sqcap$}$\sqcup$}}
\def\qed{\ifmmode\squareforqed\else{\unskip\nobreak\hfil
		\penalty50\hskip1em\null\nobreak\hfil\squareforqed
		\parfillskip=0pt\finalhyphendemerits=0\endgraf}\fi}

\begin{document}

\title{Leveraging the Power of Prediction: Predictive Service Placement for Latency-Sensitive Mobile Edge Computing}

\author{Huirong~Ma, Zhi~Zhou,~\IEEEmembership{ Member,~IEEE}, and~Xu~Chen,~\IEEEmembership{Member,~IEEE}
	\thanks{Part of the results of this journal version paper was presented at IEEE/CIC ICCC \cite{HUIRONG}.}
	\thanks{H. Ma, Z. Zhou and X. Chen are with the School of Data and Computer Science, Sun Yat-sen University, China (e-mail: mahr3@mail2.sysu.edu.cn, \{zhouzhi9, chenxu35\}@mail.sysu.edu.cn)

%(Corresponding author: Xu Chen).

%This work was supported in part by the National Key Research and Development Program of
%China under grant No.2017YFB1001703; the National Science Foundation of China (No. 61972432); the Program for Guangdong Introducing Innovative and Entrepreneurial Teams (No.2017ZT07X355);the Pearl River Talent Recruitment Program (No.2017GC010465).
}
	}

\maketitle
\renewcommand{\thefootnote}{\fnsymbol{footnote}}

\begin{abstract}
Mobile edge computing (MEC) is emerging to support delay-sensitive 5G applications at the edge of mobile networks.
When a user moves erratically among multiple MEC nodes, the challenge of how to dynamically migrate its service to maintain service performance (i.e., user-perceived latency) arises.
However, frequent service migration can significantly increase operational cost, incurring the conflict between improving performance and reducing cost.
To address these mis-aligned objectives, this paper studies the performance optimization of mobile edge service placement under the constraint of long-term cost budget.
It is challenging because the budget involves the future uncertain information (e.g., user mobility).
To overcome this difficulty, we devote to leveraging the power of prediction
and advocate  predictive service placement with predicted near-future information.
By using two-timescale Lyapunov optimization method, we propose a $T$-slot predictive service placement (\textbf{PSP}) algorithm to incorporate the prediction of user mobility based on a frame-based design.
We characterize the performance bounds of \textbf{PSP} in terms of cost-delay trade-off theoretically.
Furthermore, we  propose a new weight adjustment scheme for the queue in each frame named \textbf{PSP-WU} to exploit the historical queue information, which greatly reduces the length of queue while improving the quality of user-perceived latency.
Rigorous theoretical analysis and extensive evaluations using realistic data traces demonstrate the superior performance of the proposed predictive schemes.
\end{abstract}	


\IEEEpeerreviewmaketitle
\section{INTRODUCTION}
With the explosive growth of mobile devices, our daily life is increasingly exposed to an overabundance of mobile applications in recent years. Applications with low latency requirements, such as smart applications for vision, hearing, or mobility-impaired users, online gaming, augmented reality, and tactile computing can not be well satisfied by the cloud computing paradigm due to the long transmission latency over the Internet.
To fulfill the stringent delay requirements (typically tens of milliseconds \cite{tens}), mobile edge computing (MEC)\cite{l1,l2,l3} is proposed as a new computing paradigm to serve these applications at the edge of mobile networks, at small server clusters referred to as cloudlets \cite{l42}, fog \cite{l43}, follow me cloud \cite{l44}, or micro clouds \cite{l9}.
With the help of MEC, cloud computing and storage capabilities are moved from the core of the network to the network edge, being closer to mobile devices and users. This trend is expected to continue to be unabated and to play an important role in the next generation 5G networks for supporting latency-sensitive services \cite{5G2}.

Here an MEC node is typically a micro-data center that can host computing services, attached to a base station (or an access point), and serves the nearby devices. In the MEC paradigm, end-to-end latency is significantly reduced because it is the nearby MEC nodes instead of the remote cloud that provide computing and storage capabilities to mobile users \cite{li2018edge}. In addition, the density of 5G BSs has been increasing and is highly anticipated to reach up to 50 BSs per $km^2$ in future 5G cellular networks \cite{5G}. However, a new problem arises in this dense cellular networks due to the erratic user mobility: whether the service should be dynamically migrated among multiple MEC nodes to maintain service performance (i.e., user-perceived latency) or not.

In the paradigm of MEC, for each user, there is a computing delay associated with running the service profile in an MEC node, and there is a communication delay associated with communicating with the MEC node, and the perceived latency is determined by both computing delay and communicating delay, similar to many existing works such as \cite{Xu2018Efficient}, \cite{Xu2016Efficient}.
As an illustrative example shown in Fig. 1, a mobile user is within the service scope of the top MEC node. Obviously if we want to minimize the user perceived latency, the service should be served by the nearest MEC node, i.e., the top MEC node.
Considering user mobility, we assume in the next time slot, the mobile user would move to the service scope of the left MEC node. If the user's service is kept maintained on the top MEC node, user perceived latency will be seriously deteriorated due to the prolonged network distance.
This example shows that in order to optimize the user's experience, the mobile user's service needs to be dynamically relocated among multiple MEC nodes to follow the user mobility. And recent empirical measurement studies \cite{handover1}, \cite{handover2} have shown that, in the 5G scenario, the handover delay can be reduced and pushed to as low as 20 ms, which is greatly slighter than the communication and computing latency, and thus can be omitted in our problem. Therefore the impact of user mobility on service latency is enlarged.
\begin{figure}[t]
	\centering
	\includegraphics[width=3.5in,height=5in,clip,keepaspectratio]{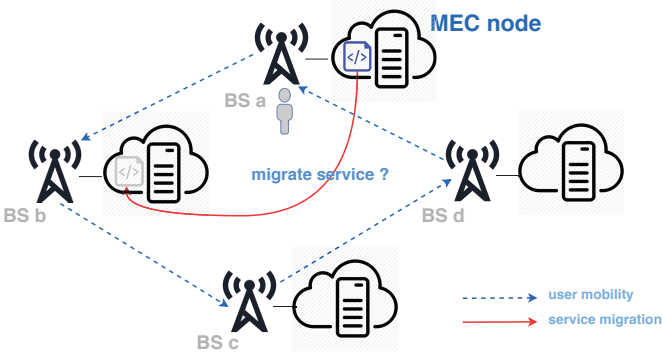}
	\caption{An example of dynamic service placement when a user roams through several locations}
	\label{fig:label}
\end{figure}

Unfortunately, the dynamic service placement problem is non-trivial. On the one hand, if the user's service is aggressively placed on the nearest MEC node, it may cause huge migration cost (e.g., the bandwidth and power consumption). On the other hand, if the user's service doesn't follow the user mobility, the user's experience dictated by perceived latency may be poor. Therefore, there is a trade-off between migration cost minimization and user-perceived latency minimization. In reality, user-perceived latency minimization often conflicts with migration cost minimization. Thus an efficient dynamic service placement strategy should carefully weigh performance-cost trade-off in a cost-efficient manner. In this paper, we consider a long-term migration budget (which is pre-defined yearly or monthly by the operators in practice) for minimizing the user perceived latency over a long run. Therefore, a challenge issue arises: with limited knowledge of user mobility, how can a operator make dynamic service placement decisions to minimize user perceived latency, while maintaining long term migration budget?

To cope with the challenge issue, we will make dynamic service placement decisions under the long-term migration budget constraint and develop effective dynamic service placement algorithm to leveraging the power of prediction. The main contributions of this paper are summarized as follows:
\begin{itemize}
	%\item By applying Lyapunov optimization technique \cite{l7}, we first design a one-slot reactive online service placement algorithm (\textbf{OSP}) to decompose the long-term optimization problem into a series of real-time optimization problems without requiring future information (such as user mobility). %Our approach is tunable by a control parameter $V$ which indicates the relative importance of migration cost minimization versus service performance (i.e., user perceived latency). The \textbf{OSP} algorithm yields a time-averaged user perceived latency within $O(1/V)$ bound of the optimal value, and guarantee an $O(V)$ long-term migration budget.
	\item Inspired by the recent advances in predictive service placement \cite{l9}, we devote to leveraging the power of user mobility prediction to find the optimal service placement decisions. By using two-timescale Lyapunov optimization method \cite{l7}, we propose a $T$-slot predictive service placement (\textbf{PSP}) algorithm and incorporate the prediction of user mobility in our approach properly.  We show that this problem is equivalent to a shortest-path problem which can be readily solved.  %With an accurate predication of user mobility in a short future, the operator is able to make more efficient and proactive service placement decisions.
	\item In \textbf{PSP}, we design the algorithm only based on the current and predicted future information, without the information for previous frames. Thus, as an enhancement, we further exploit the historical queue information to add a new weight adjustment scheme for queue management named weight-update method for \textbf{PSP} algorithm (i.e., \textbf{PSP-WU}), which greatly reduces the length of queue while improving the quality of user-perceived latency.
	\item Extensive experiments based on real-world data traces are carried out to verify the theoretical analysis. Moreover, the effectiveness of the proposed algorithms is demonstrated by comparisons with 4 benchmark algorithms including one predictive lazy migration algorithm. It is shown that \textbf{PSP} scheme is highly efficient and achieves superior performance in user perceived latency than the commonly-adopted online optimization mechanisms without prediction, and \textbf{PSP-WU} further reduces the user perceived latency at the same migration cost budget over the $\textbf{PSP}$ scheme.
	%Extensive experiments based on real-world data traces are carried out to evaluate the performance of the proposed algorithms, which demonstrate that the proposed \textbf{PSP} scheme is highly efficient and \textcolor{blue}{achieves superior performance in} user-perceived latency than the commonly-adopted online optimization mechanisms without prediction, and \textbf{PSP-WU} further reduces the user-perceived latency at the same migration cost budget over the $\textbf{PSP}$ scheme.
\end{itemize}

The rest of this paper is organized as follows. We introduce the corresponding related work in Section II and present the problem formulation in Section III in detail. In Section IV, the benchmarking online service placement (\textbf{OSP}) algorithm is presented. Section V proposes the $T$-slot predictive service placement (\textbf{PSP}) algorithm based on the two-timescale Lyapunov optimization method. Section VI proposes the enhanced \textbf{PSP-WU} algorithm to use the historical queue information to present the value of the queue in each frame. Performance evaluation is carried out in Section VII and Section VIII concludes this paper.

\section{RELATED WORK}
In recent years, mobile edge computing has attracted more and more attention \cite{l112,l113,l114,zhou2019edge}, especially the service placement problem \cite{l15}. When user mobility exists, it is necessary to consider dynamic service migration. For example, it can be beneficial to migrate the service to an MEC node closer to user. Tracking users and devices mobility is a key challenge in implementing efficient service placement in MEC. 
Essentially, existing researches about user mobility of service placement can be classified into the following three categories: i) totally unpredictable user mobility, ii) prefect prediction of user mobility, and iii) limited prediction of user mobility.

Many researches have devoted to the consequences of unpredictable user mobility, such as \cite{l24,l19,l20,ml1}. For example, the cloud service providers can obtain users' location from mobility management entity is assumed in \cite{l24}. The main approach in \cite{l19,l20} is to addresses the trade-off of cost between user perceived quality and migration cost by modelling the service migration procedure as a Markov Decision Process (MDP). \cite{l20} shows when the mobile user follows an one-dimensional asymmetric random walk mobility model, the optimal policy for that service migration is a threshold policy. Such a model is suitable where the user mobility follows a Markov chain model.
However, the Markovian assumption only works for a specific class for cost functions \cite{ml1}, and is not always valid in all cases \cite{l199}. 

Another stream of recent work resorts to a stronger assumption of prediction on future users' mobility. 
For example, authors of \cite{ml2} handle the system dynamic nature by first predicting the user's movement, and then exploiting this prediction for dynamic VM placement before the User Equipment (UE) starts offloading. However, \cite{ml2} only tries to reduce latency by reducing the communication delay between the user and the Virtualized Edge (VE), ignoring the computational delay. A pronounced difference is that our work considers the user perceived latency is determined by both computing delay and communication delay, and thus our algorithm can be more practical.
The authors in \cite{l9} consider the case in which an underlying mechanism is used to predict the future costs of service hosting and migration, and the prediction error is assumed to be bounded. However, these predictions are hard to be computed accurately in real environments, and the inherent network dynamics may further escalates this difficulty.
A user-centric location prediction approach and a factor graph learning model is proposed in \cite{l100}, and its algorithm utilize one-shot prediction of next location to help adaptive service migration decision-making to personalized service migration in mobile edge computing (MEC) to show the benefit of precise location prediction. In comparison, our predictive service placement strategy looks ahead over a time frame (contains multiple time slots) based on the prediction and the influence of prediction accuracy to performance is considered. 

With the development of machine learning \cite{machine}, it is possible to predict the future mobility of users, and the accuracy of the prediction is getting higher. 
In our study, we devote to the situation of limited prediction of user mobility.
A closely related work \cite{l11} applies one-slot Lyapunov optimization technique and develop an approximation algorithm based on the Markov approximation technique to approach a near-optimal solution.
It is worth noting that our work substantially differs from and complements to \cite{l11} in the following two aspects: i) we incorporate the prediction of user mobility; ii) to design efficient algorithm, the work \cite{l11} is based on one-slot Lyapunov optimization without prediction and Markov approximation technique, while we take the advantages of one-slot Lyapunov optimization technique, two-timescale $T$-slot Lyapunov optimization technique and Dijkstra algorithm into consideration, and our algorithm achieves better performance.

This work significantly extends the preliminary work \cite{HUIRONG}. Aiming at a paradigm shifting from reactive to predictive and better illustrate the performance of \textbf{PSP}, we design a benchmark online service placement algorithm named one-slot reactive online service placement  (\textbf{OSP}) algorithm without requiring future information (such as user mobility).
Rather than using only the queue information at the current frame and being memoryless of the past frames' information, we further exploit the historical queue information to add a new weight adjustment scheme for queue management named weight-update method for \textbf{PSP} algorithm (i.e., \textbf{PSP-WU}), which greatly reduces the length of queue while improving the quality of user-perceived latency.  We have also provided new theoretical results on the performance analysis of  online service placement without prediction algorithm and predictive service placement algorithm. Besides the experimental results in the ICCC paper, in this paper we add more performance comparison with different prediction methods, different prediction window size $T$ and different average migration cost to verify the efficacy of our proposed schemes. 
\begin{table}[t]
	\caption{Parameter table of system model}
	\begin{center}
		\renewcommand\arraystretch{1.5}
		\begin{tabular}{|p{0.82cm}||p{7cm}|}
			\hline
			Notation & \multicolumn{1}{c|}{Definition} \\
			\hline
			$\mathcal{N}$ & set of MEC nodes\\
			\hline
			$t$, $\tau$ & index of time slot\\
			\hline
			$\widetilde{T}$ & index of the total time slots \\
			\hline
			$W$ & index of prediction window size\\
			\hline
			$T$ & index of one frame size\\
			\hline
			$i$, $j$ & index of location\\
			\hline
			$x_i(t)$ & whether the service is placed at MEC node $i$ (=1) or not (=0) \\
			\hline
			$H_i(t)$ & the service latency from user to location $i$\\
			\hline
			$L(t)$ &  user-perceived latency\\
			\hline
			\multirow{1}*{$P_{ji}(t)$}
			& the cost of migration the service from source location $j$ to destination location $i$\\
			\hline
			$E(t)$ & total service migration cost\\
			\hline
			\multirow{1}*{$E_{avg}$} & the long-term time-averaged cost budget over a time span of $T$ time slots\\
			\hline
			$V$ & Lyapunov control parameter\\
			\hline
			$\eta$ & a positive controllable parameter \\
			\hline
			$\theta$ & a positive controllable parameter, and $\in(0,\eta]$ \\
			\hline
			$\beta$ & a positive controllable parameter, and $\in[0,1]$\\
			\hline
		\end{tabular}
	\end{center}
\end{table}
\section{PROBLEM FORMULATION}
As shown in Fig. 1, we consider a scenario where a user moves frequently in several locations and uses latency-sensitive service which require real-time analysis, e.g., data streaming analysis, and the user sends continuous requests to the network operator at each time slot, thus has higher requirements on service quality. In our study, a network operator runs a set $\mathcal{N}=\{1,2,..,N\}$ of MEC nodes. Each MEC node is attached to a base station (or an access point), through high-speed local area network (LAN).
We consider that the user is mobile and always associated with nearest MEC node, and then through network connections the user access to its service VM or container at the service hosted MEC node.
In general, the models of user mobility can be classified into following types, i.e., the random walk model \cite{m1}, the random waypoint model \cite{m2}, the fluid flow model \cite{m3}, the Gauss-Markov model \cite{m4} and measured from real data \cite{m5} etc. In our study, we consider the user mobility measured from real world data trace.

Without loss of generality, we assume that the system operates in a time-slotted structure within a large time span which helps to capture the dynamic of user mobility, and its timeline is discretized into time slots $t\in\mathcal{T} =\{0,1,2,...,\widetilde{T}\}$.
At each time slot $t$, the mobile user sends its MEC service requests to the network operator, which then determines whether to migrate the service (i.e., the user's subscribed MEC container or VM service with a fixed resource capacity) or not. By carefully navigating the performance-cost trade-off, the operator will decide where to migrate the service (among the multiple MEC nodes). Note that in this study we take a user-centric point of view in order to provide personalized and fine-grained service placement for each individual user. We will further consider the system-wide predictive network management in a future work. 
Notations in the paper are listed in Table I for ease to reference.

\subsection{Service Placement Model}
In order to satisfy user's requirement of Quality-of-Service (QoS) (i.e., low service latency), the service should be dynamically migrated among multiple MEC nodes to track the user mobility. 
To denote the dynamic service placement decision variable, we use a binary indicator $x_i(t)$, if the service is placed at the MEC node $i\in \mathcal{N}$ at time slot $t$, we let $x_i(t)= 1$, otherwise $x_i(t)= 0$.
At a given time slot, the service is placed at one and only one MEC node, and the constraints for $x_i(t)$ are as follows.
\begin{equation}
\sum_{i=1}^N x_i(t)=1,   \forall t,   
\end{equation}
\begin{equation}
x_i(t)\in\{0,1\}.    
\end{equation}
With the service placement decision defined above, next we start to formulate the user perceived latency determined by the service placement decision $x_i(t)$.

\subsection{QoS Model}
In the paradigm of MEC, the user's QoS is determined by both computing delay and communicating delay.
In general, the computing delay is determined by user's current service request (i.e., the required amount of CPU cycles) and the communication delay depends on the network transmission latency between the user and the MEC node hosting its service. 
Without loss of generality, given the service request information and the current location of user, we use a general term $H_i(t)$ to denote the service latency (i.e., computing delay and communicating delay) from the user to its service hosted MEC node $i$ at the time slot $t$. Thus we capture the fact that the users task can vary from time to time and the resources at different MEC nodes for handling the users task can also fluctuate dynamically. 
For instance, let $z(t)$ denotes the nearest MEC node that the user associated with at time $t$. And the tuple $<I(t), K(t)>$ denote the profiles of the user service task at time $t$, where $I(t)$ denotes the input data volume, $K(t)$ denotes the computation workloads (CPU cycles) for data processing. Then the total service latency $H_i(t)=\frac{I(t)}{B_{z(t)}}+\frac{I(t)}{B_{z(t),i}}+\frac{K(t)}{D_i(t)}$, where $B_{z(t)}$, $B_{z(t),i}$ and $D_i(t)$ denote the network bandwidths between user and its associated MEC node and between user's associated MEC node and its service hosted MEC node, and the current computing power of its service hosted MEC node, respectively.
Since the service is placed at one and only one MEC node at each time slot $t$, the user's service latency can be expressed by:
\begin{equation}
L(t) = \sum_{i=1}^{N}x_i(t)H_i(t).
\end{equation}
\subsection{Migration Cost Model}
Migrating service among multiple MEC nodes dynamically incurs additional operational cost.
Especially, transferring the service across edges will cause modest usage of scarce and expensive wide area network (WAN) bandwidth and increase the energy consumption of network devices (such as routers and switches). To model the operational cost of cross-edge service migration, we use $P_{ji}(t)$ to denote the cost of migrating the service from source MEC node $j$ to destination MEC node $i$. For instance, we can denote $S(t)$ as the current data size of the virtual machine or container that hosts the user's service. Then the migration cost $P_{ji}(t)=S(t)B_{ji}(t)$, where $B_{ji}(t)$ denotes the unit network bandwidth cost for service migration. Also, other cost terms such as energy cost for service migration can be also added into the equation above if needed.
And we assume $P_{ji}(t)=0$,  $\forall j = i $. Given the service placement decision $x_j(t-1)$ at time slot $t-1$, and $x_i(t)$ at time slot $t$, the service migration cost of user at time slot $t$ can be further denoted as: $E(t) = \sum_{i=1}^N \sum_{j=1}^NP_{ji}(t)x_j(t-1)x_i(t)$.

\subsection{Problem Formulation}
Assigning different weights to the conflicting objectives and  optimizing the sum of them is the most commonly method to optimize the multiple conflicting objectives in a balanced manner such as \cite{l12}, \cite{l122}. However, in our problem, it is not simple to correctly define the weights of performance and cost in the real world. In addition, we take the fact that the operator is generally sensitive to the operational cost, and hence define a long-term migration cost budget (which can be specified by the operator as per the factors such as the operational cost and the user's service level) to minimize the user-perceived latency over the long run.
We introduce $E_{avg}$ to denote the long-term time-averaged cost budget over a time span of $T$ time slots, which satisfies: 
\begin{equation}
\lim_{\widetilde{T}\to\infty} \cfrac{1}{\widetilde{T}}\sum_{t=0}^{\widetilde{T}-1}E(t)\le  E_{avg}.
\end{equation}
Therefore, our problem is to minimize the long-term time-average service delay under the long-term cost budget constraints, which equals to a stochastic optimization problem as follows:
\begin{equation}
\begin{aligned}
&\min \lim_{\widetilde{T}\to\infty} \cfrac{1}{\widetilde{T}} \small \sum_{t=0}^{\widetilde{T}-1}\{L(t)\} \\ 
&\qquad s.t. (1)-(4). 
\end{aligned}
\end{equation}

Generally speaking, one of the main challenges hindering the promotion of the optimal long-term policy of (5) is that it needs future information (i.e. user mobility and other information). Under future information, the operator can make the global service placement decision for simultaneously minimizing the long-term time-averaged QoS and enforcing the time-averaged cost budget. Although the long-term service placement optimization problem is decomposed into real-time decoupling problem, it is of great importance to prevent frequent service migration by using long-term migration cost constraint.

Fortunately, the long-term migration budget constraint (4) in this optimization problem can be regarded as the queue stability control, i.e., $\lim_{\widetilde{T}\to\infty}\frac{\small 1}{\small \widetilde{T}}\sum_{t=0}^{\widetilde{T}-1}E(t) \leq E_{avg}$.
By ensuring that the time-averaged migration cost is lower than the long-term budget, we get the stable queue. In the following sections, we will elaborate how to solve the problem in the settings without and with prediction, respectively.

\newtheorem{theorem}{Theorem}
\newtheorem{lemma}{Lemma}
\section{BENCHMARK: ONLINE SERVICE PLACEMENT WITHOUT PREDICTION}
In this section, as a key benchmark and foundation for predictive service placement, we first describe the traditional Lyapunov optimization-based online service placement without prediction \textbf(OSP) algorithm. To solve (5), we will introduce some key definitions.

\subsection{Problem Transformation via Lyapunov Optimization}
Considering the dynamics and randomness of the system (time-varying and uncertainly user mobility), the main challenge of (5) is to balance the performance-cost trade-off in a cost-efficient manner without global information in the long run. By using Lyapunov optimization to introduce a virtual queue which measures the long-term budget, a desirable balance between user perceived latency and migration cost can be achieved, while maintaining the migration cost queue stable.
We define a virtual queue, and $Q(t)$ is the queue length at time slot $t$, which represents the historical measurement of the migration cost at time slot $t$. %We assume the queue is initially empty (i.e., $Q(0) = 0$). 
The queue length evolves according to the migration cost and $ E_{avg}$ as:
\begin{equation}
Q(t+1) = \max[Q(t) + E(t) - E_{avg},0].
\end{equation}
Intuitively, the value of $Q(t)$ can be used as an evaluation criterion to evaluate the condition of migration cost.
A large value of $Q(t)$ implies the cost has exceeded the long-term budget $E_{avg}$. To guarantee the time-averaged service migration cost is under the value of $E_{avg}$ (i.e., Eq. (4)), we must keep the queue stable, i.e., $\lim_{\widetilde{T}\to\infty}\mathbb E\{Q(\widetilde{T})\} / \widetilde{T} = 0$. By totally summing and rearranging the inequality $Q(t+1)\ge Q(t) +E(t) - E_{avg} $ derived from Eq. (6) over time slot $t\in\{0,1,2,...,\widetilde{T}-1\}$, we have:
$$\cfrac{Q(\widetilde{T})-Q(0)}{\widetilde{T}} +  E_{avg} \ge \cfrac{1}{\widetilde{T}}\sum_{t=0}^{\widetilde{T}-1}E(t).$$
By initializing $ Q(0) = 0 $, and taking expectations of the above inequality we can gain:
$$\lim_{\widetilde{T}\to\infty} \cfrac {\mathbb E\{Q(\widetilde{T})\}}{\widetilde{T}} + E_{avg} \ge \cfrac{1}{\widetilde{T}}\sum_{t=0}^{\widetilde{T}-1}\mathbb E\{E(t)\}.$$
Therefore, the stability of the virtual queue (i.e., $Q(\widetilde{T})<\infty$) ensures that the time-averaged migration cost is beneath the long budget $E_{avg}$.

In order to stabilize the virtual queue (i.e., to ensure $Q(\widetilde{T})<\infty$), the quadratic Lyapunov function is defined as follows:
\begin{equation}
L(\Theta(t))\triangleq \cfrac{1}{2}Q(t)^2.
\end{equation}
In general (7) is a measurement of overruncost level \cite{l7} in queue. For example, if the value of $L(\Theta(t))$ is small, the queue backlog is small too. This guarantees a strong stability of virtual queue.
The core idea of solving original constrained stochastic optimization problem in (5) is to minimize the Lyapunov drift-plus-penalty function which incorporates  queue stability and user-perceived latency jointly. 
\begin{algorithm}[tp]
	\caption{Algorithm for Online Service Placement without Prediction (OSP)}
	\begin{algorithmic}[1]
		\REQUIRE  ~~\\  
		We set the cost queue backlog $Q(0) = 0$ at beginning.
		\ENSURE ~~\\ 
		\FOR {each time slot $t \in \{0,1,2,...,\widetilde{T}-1\}$}
		\STATE $\mathbf{Solve}$ the problem (11): $x_i(t) = \arg \min(VH_{i}(t)+\rho_i(t))$.
		\STATE $\mathbf{Update}$ the virtual queue: run (6) based on policy $x^{*}(t)$.
		\ENDFOR
	\end{algorithmic}
\end{algorithm}
\subsection{Joint Queue Stability and User-Perceived Latency Minimization}
We use \emph{one-step  conditional Lyapunov  drift} to explain the virtual queue stability as follows:
\begin{equation}
\Delta{\Theta(t)} \triangleq \mathbb E{[L(\Theta({t+1})-L(\Theta(t)) | \Theta(t)]}. 
\end{equation}

According to the Lyapunov drift theorem \cite{l7}, if Lyapunov drift is small (i.e., do one's best to push the migration cost under long-term budget constraint), then the virtual queue is stable. It indicates the long-term time-averaged migration cost could be less than or equal to $E_{avg}$ which has been predefined as cost budget.

We intend to seek out a current service placement policy thus the perceived latency and migration cost can be balanced. Incorporating queue stability into delay performance, a \emph{Lyapunov drift-plus-penalty} function can be defined to solve the real-time problem as follows:
\begin{equation}
\Delta(\Theta(t)) + VL(t) ,
\end{equation}
where \emph{V} is a non-negative control parameter. By using \emph{V}, we can adjust the trade-off between latency performance and migration cost queue backlogs, and we pay more attention to delay performance compared with migration cost budget. 
Moreover, the \emph{drift-plus-penalty} functions' performance guarantee is shown in the following lemma.
\begin{lemma}
	In each time slot $t$, given the service placement decision $x_i(t)$, the following statement holds:
	\begin{equation}
	\begin{split}
	&\Delta(\Theta(t)) + V\mathbb E{[L(t)| \Theta(t)]} \\
	&\leq B +V\mathbb E{[L(t)| \Theta(t)]}+ \mathbb E{\bigr \{Q(t)[E(t)-E_{avg}|\Theta(t)\bigr]\bigr\}} , 
	\end{split}    
	\end{equation}
	where $B=\cfrac{1}{2}(E_{avg}^2+E_{\max}^2)$ is a constant value, and $E_{\max} = \max_tE(t)$.
\end{lemma}
The proof of Lemma 1 is given in our online technical report \cite{Appendix}. 
Based on Lemma 1, the \emph{drift-plus-penalty} function yields an upper bound at every time slot.

\subsection {Algorithm for Online Service Placement without Prediction}
In this section, we devote to minimizing a series of real-time \emph{drift-plus-penalty} supremum bounds which are converted from (9). Due to the $\max[*]$ terms in Eq. (6), the minimization of \emph{drift-plus-penalty} expression in Eq. (9) begins to thicken. Based on Lemma 1, we seek to minimize the right side of Eq. (10). Thus, rearranging it for a concise form, we have the following optimization problem:
\begin{equation}
\begin{aligned}
\min & \ \sum_{i=1}^{N}x_i(t)[VH_i(t)+\rho_i(t)]  \\
& \quad \text{s.t.} \quad (1)-(3),
\end{aligned}     
\end{equation}
where $\rho_i(t)=Q(t)\sum_{j=1}^{N}x_j(t-1)P_{ji}(t)$ is a constant value at every time slot $t$.

According to Algorithm 1, $x_i(t)= 1$, if $i = i^{*}$, otherwise $x_i(t)= 0$. Fortunately, this real-time optimization problem can be solved easily, by comparing the value of $VH_{i}(t)$+$\rho_i(t)$ for $i \in\{1,2,...,N\}$. After comparing $N-1$ times, we get the minimum of (11) and return the index of the location, so that we get the policy $x^{*}(t)$, which determines where to place the service at each time slot $t$.

\subsection{ Performance Analysis of \textbf{OSP}}
For the ease of exposition, we analyze the performance of \textbf{OSP} algorithm by assuming that the user perceived latency $H_i(t)$ is independent and identically distributed (i.i.d.) over time slots. Let $\mathbb E\{{E^{+}(t)}\}$ be the the migration cost by the stationary service placement policy. The following assumption ensures that there exists a service placement algorithm that satisfies the stability constraint: $\lim_{\widetilde{T}\to\infty}\sup \cfrac{1}{\widetilde{T}} \sum_{t=0}^{\widetilde{T}-1}\mathbb E\{{Q(t)}\}<\infty$.  According to \cite{l7}, we assume that the current migration cost holds: 
\begin{equation}
\mathbb E\{E^{+}(t)\} -{E_{avg}} \leq - \eta ,
\end{equation} 
where $\eta>0$ is a finite constant that represents the distance between the time-averaged migration cost by some control policy and long-term cost budget.

We define $L_{av}^{OSP}$ and $Q_{av}^{OSP}$ as the long-term average user perceived latency and average queue length of \textbf{OSP} algorithm, respectively. Theorem 1 establishes the upper bounds of $L_{av}^{OSP}$ and $Q_{av}^{OSP}$.
\begin{theorem}For any non-negative control parameter $V$, the long-term user-perceived latency and cost budget by proposed \textbf{OSP} algorithm satisfies:
	\begin{equation}
	L_{av}^{OSP}  \triangleq \lim_{\widetilde{T}\to\infty}\sup\cfrac{1}{\widetilde{T}}\sum_{t=0}^{\widetilde{T}-1}\mathbb E\{{L}(t)\} \leq L_{av}^{*}+\cfrac{B}{V},
	\end{equation}
	\begin{equation}
	Q_{av}^{OSP} \triangleq  \lim_{\widetilde{T}\to\infty}\sup\cfrac{1}{\widetilde{T}}\sum_{t=0}^{\widetilde{T}-1}\mathbb E\{Q(t)\} \leq \cfrac{B+VL_{\max}}{\eta},
	\end{equation}
\end{theorem}
where $B=\cfrac{1}{2}(E_{avg}^2+E_{\max}^2)$ is a constant value, $\eta$ is a positive parameter, $L_{av}^{*}$ is the optimal expected time averaged latency of problem (5), and $L_{\max}=\max_tL(t)$.

The detailed proof of Theorem 1 is given in our online technical report \cite{Appendix}.
Theorem 1 demonstrates an $[O(1/V), O(V)]$ service performance and cost budget trade-off achieved by \textbf{OSP} algorithm. Nevertheless, we should emphasize that the \textbf{OSP} algorithm only uses the current information for service placement decision making, and in order to satisfy the long-term cost budget constraint, \textbf{OSP} algorithm would tend to make conservative placement decisions and it is hard to further improve the service performance. Thus, we advocate to leverage the power of prediction and devise a predictive service placement algorithm to make informed decisions and achieve proactive service placement for service performance enhancement. 
\section{PREDICTIVE SERVICE PLACEMENT}
In this part, we elaborate the predictive service placement algorithm design in details. As mentioned earlier, with the development of machine learning (e.g., deep learning) \cite{machine}, it is now possible to accurately predict user's service pattern (e.g., mobility) in a short future.
In order to well integrate the predicted information, we resort to the two-timescale Lyapunov optimization framework and propose a $T$-slot predictive service placement (\textbf{PSP}) algorithm based on a frame manner. Then, we translate $T$-slot predictive service placement problem into a shortest-path problem which can be readily solved based on the predicted near-future information.
\begin{figure}[t]
	\centering
	\includegraphics[width=3.5in,height=4.85in,clip,keepaspectratio]{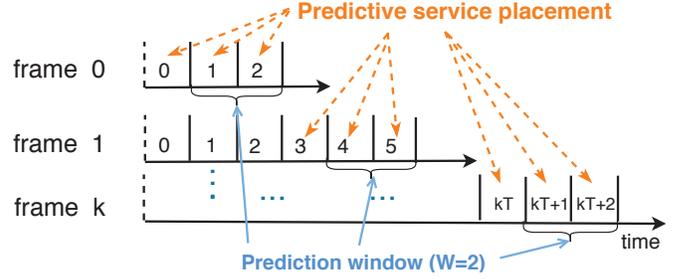}
	\caption{ Frame-based structure for predictive service placement }	
\end{figure}
\subsection{Frame-based Predictive Service Placement Mechanism}
The key idea of predictive service placement algorithm design is to look ahead over a time frame based on the prediction and then plan the predictive service placement accordingly. Such predictive decision makings are repeated every time frame. Specifically, for a frame-based structure, as shown in Fig. 2, the time interval is the $k$-th ($k\in\{0,1,2,...\}$) frame which contains time slots $kT,kT+1,...,kT+T-1$, and the length of each frame is $T$.
We assume that at time slot $t=kT$, i.e. the beginning of the current $k$-th frame, the operator will consider all the locations of the user in the entire frame predicted by advanced machine learning methods such as LSTM \cite{l27} with a prediction window size $W=T-1$.
In our prediction process, we adopt a simple all-connected output window position information to predict the user mobility. Since the prediction error ascents dramatically as the number of time slots grows, the location in the earlier slots can be predicted with a higher accuracy.
With the predicted user mobility in a whole frame, the operator makes the optimal predictive service placement decisions $\{x_i(\tau)\}$, $\tau \in \mathcal{T}_k$, for the entire $k$-th frame.

By using the prediction information over the time frames, we can achieve proactive and more informed service placement decisions. However, the key challenge is to satisfy the long term migration cost budget constraint in original problem (5). This is not easy to tackle since the predictive decisions are made over a larger time-scale of frames, while the migration cost occurs over a smaller time-scale of time slots. To address this issue, we hence propose a two-timescale Lyapunov optimization framework.

\subsection{Problem Transformation via Two-timescale Lyapunov Optimization }
Similar to \textbf{OSP} algorithm, for \textbf{PSP} algorithm, the queue function and the Lyapunov function at each time slot $t$ are the same as in Eq. (6) and Eq. (7). By using the two-timescale conditional Lyapunov drift to interpret the virtual queue stability, we define a frame based $T$-slot Lyapunov drift, $ \Delta{_T(t)} $ as the expected change in the Lyapunov function over $T$ slots:
\begin{equation}
\Delta{_T(t)} \triangleq \mathbb E{[L(\Theta({t+T})-L(\Theta(t)) | \Theta(t)]}.
\end{equation}
Following the same idea of traditional Lyapunov optimization approach, we add the expected user-perceived latency over a frame of  $T$ slots (i.e., a penalty function, $\mathbb E{\{\sum_{\tau=t}^{t+T-1}L(\tau)\}}$) to (15) to obtain the \emph{drift-plus-penalty} term for the $k$-th frame as:
\begin{equation*}
	\Delta{_T(kT)} + V\mathbb E{\bigr\{\sum_{\tau=kT}^{kT+T-1}L(\tau)| \Theta(kT)\bigr\}}.
\end{equation*}

The following Lemma characterizes the upper bound for the \emph{drift-plus-penalty} term.
\begin{lemma}
For any value of $Q(kT)$ and service placement policy {$x^*(\tau)$}, $\tau \in \mathcal{T}_k$, the "drift-plus-penalty" term for the $k$-th frame satisfies:
\begin{equation}
\begin{split}
&\Delta{_T(kT)} + V\mathbb E{\bigr\{\sum_{\tau=kT}^{kT+T-1}L(\tau)| \Theta(kT)\bigr\}} \\& \leq BT
+\mathbb E{\biggr\{\sum_{\tau=kT}^{kT+T-1}Q(\tau)[E(\tau)-E_{avg}]|\Theta(kT)\biggr\}}\\&
+V\mathbb E{\biggr\{\sum_{\tau=kT}^{kT+T-1}L(\tau)|\Theta(kT)\biggr\}},
\end{split}
\end{equation}
	where $B$ is the constant defined in Lemma 1, \emph{V} is a non-negative control parameter, and $E_{\max} = \max_tE(t)$. The proof of Lemma 2 is given in our online technical report \cite{Appendix}.
\end{lemma}

Note that minimizing the drift-plus-penalty function in (16) requires future knowledge of user mobility and migration cost at each time slot (which can be easily obtained through machine learning based prediction and system measure) as well as the queue backlog information over a time frame. However, for the queue backlogs, considering the frame based structure and their accumulative nature over time slots, it is hard to well predict such information, thus the prediction errors should be considered. Hence, we propose to take an approximation by setting the future queue backlogs as their current backlogs at slot $t=kT$, as given by
\begin{equation}
	\hat{Q}(\tau)=Q(t), \quad\forall \tau=t,...,t+T-1,
\end{equation}
where $\hat{Q}(\tau)$ is the approximation queue backlogs.

Taking approximation of (17) to be invariant in the coming time window, (17) reduces the complexity while suits more on our frame-based design.
Based on (17), we establish the following lemma:
\begin{lemma}
		At any time slot $\tau$, the differences between the approximated and actual queue backlogs in (16) are bounded by
		\begin{equation}
		|Q(\tau)-\hat{Q}(\tau)|\leq Tw_Q,
		\end{equation}
\end{lemma}
where the constant $w_Q=\max\{E_{avg},E_{\max}\}$.

\emph{Proof.} For any two consecutive slots $\tau$ and slot $\tau+1$, the difference of the queue backlogs is bounded, i.e., $|Q(\tau+1)-Q(\tau)|\leq w_Q$, where $w_Q$ is the maximum difference between the given long-term migration cost and the maximum of the migration cost, denoted by $\max\{E_{avg},E_{\max}\}$. According to (17) and the inequality $|a+b|\leq |a|+|b|$, we have $|Q(\tau)-\hat Q(\tau)|=|Q(\tau)-Q(t)|=|\sum_{t_0=t}^{\tau}[Q(t_0+1)-Q(t_0)]|\leq (\tau-t)w_Q\leq Tw_Q$, where $t=kT$ and $\tau=t,...,t+T-1$. Therefore, (18) is proved.

Based on Lemma 2 and Lemma 3, we further state the following lemma to get the following approximated drift-plus-penalty term:
\begin{lemma}
	For any value of $Q(kT)$ and service placement policy {$x^*(\tau)$}, $\tau \in \mathcal{T}_k$, the following inequality holds for any $\theta$:
	\begin{equation}
	\begin{split}
	&\Delta{_T(kT)} + V\mathbb E{\bigr\{\sum_{\tau=kT}^{kT+T-1}L(\tau)| \Theta(kT)\bigr\}} \\& \leq BT+ VTC(\theta)
	\\&-\mathbb E{\biggr\{\sum_{\tau=kT}^{kT+T-1}\theta[(k+1)T-\tau] \hat{Q}(\tau)|\Theta(kT)\biggr\}}.
	\end{split}
	\end{equation}
\end{lemma}
The detailed proof of Lemma 4 is given in our online technical report \cite{Appendix}.	

\subsection {T-slot Predictive Service Placement Algorithm}
We next present the $T$-slot predictive service placement algorithm for the optimal decision makings. The key idea is to minimize  the right side of the approximated drift-plus-penalty term in Eq. (19), and accordingly we can derive a near-optimal solution for the predictive service placement problem. Specifically, we solve the following problem in each time frame $k \in \{0,1,2,...,K-1\}$:
\begin{equation}
\begin{aligned}
\min \sum_{ \tau=kT}^{kT+T-1}\{ Q(kT)&(E(\tau)-E_{avg}+\theta[(k+1)T-\tau] )\\&+VL(\tau)\}  \\
&\text{s.t.} \quad (1)-(3).
\end{aligned}
\end{equation}
The basic idea of \textbf{PSP} is to balance the average user-perceived latency and average queue length of each frame. Besides $V$, we further introduce another positive controllable parameter $\theta$ in \textbf{PSP}, which captures the variance of queue length within each frame. The intuition is that, with $\theta[(k+1)T-\tau]$, the earlier slot's predictive location can be assigned with larger weights than those latter slots within the frame. In this way, when the other conditions are the same, user's location in the earlier slots within the frame is more important than the latter time slot's locations. This is aligned with the fact that the location in the earlier slots can be predicted with a higher accuracy. By putting more weights on time slots with more accurate prediction information, we can further improve the overall algorithm performance. We illustrate such an intuition in Section VII.
\begin{algorithm}[tp]
	\caption{Predictive service placement}
	\begin{small}
		\begin{algorithmic}[1]
			\STATE 	Define variables $i$ and $j$ to represent the possible decision respectively in the current and previous time slot.	
			\STATE Define array $\pi_{i}$ and $\zeta_{j}$ for all $i$,$j$, where $\pi_{i}$ and $\zeta_{j}$ are the optimal decisions from the beginning $t$ to the current and previous time slot when the decision is $i$ respectively.
			\STATE Define variables $\phi_{i}$ and $\varphi_{i}$ for all $i$ to count the cumulative cost from the beginning $t$ respectively to the current and previous time slot when the decision  is $i$.
			\STATE Define variables $D(i,t)$ to represent the service cost when the decision is $i$ at time slot $t$.
			\STATE  Initialize the cost queue backlog $Q(0) = 0$.
			\FOR {each frame  $k \in \{0,1,2,...,K-1\} $}
			\FOR { all $i$}
			\STATE  $\pi_{i}=\emptyset$.
			\STATE  $\phi_{i} = 0$.
			\ENDFOR
			\FOR { $t = kT,kT+1,...,kT+T-1$}
			\FOR { all $i$}
			\STATE $\zeta_{i}$ = $\pi_{i}$.
			\STATE $\varphi_{i}$ = $\phi_{i}$.
			\ENDFOR
			\FOR {all $i$}
			\STATE $j_{opt}$ = $\arg\min_{j}\{\varphi_{j}+D(i,t)\}$.
			\STATE $\pi_{i}(kT,...,t)=\zeta_{j_{opt}}\cup i$.
			\STATE $\phi_{i}=\varphi_{j_{opt}}+D(i,t)$.
			\ENDFOR
			\ENDFOR
			\STATE  $i_{opt}=\arg\min_{i}\phi_{i}$.
			\STATE  The optimal policy in each frame $k$ is: $\pi_{i_{opt}}( kT,kT+1,...,kT+T-1)$.
			\STATE $\mathbf{Update}$ the virtual queue: run (6) based on $\mathbf{x^{*}(\tau)}$ at each time slot $\tau$ in the frame $k$.
			\ENDFOR
		\end{algorithmic}
	\end{small}
	\label{fig:alg. 3}
\end{algorithm}

Although the queue-length in a frame we set as a constant value $Q(kT)$, $E(t)$ still is a time-coupling term for which we can not solve the problem (20) at each time slot. Fortunately, the problem in (20) is equivalent to a shortest-path problem and can be solved approximately by using DijKstra algorithm \cite{ldiji}.
We calculate the minimum of problem (20) to get a policy $x^*(\tau)$, and then make the decision to place the service. The \textbf{PSP} algorithm is shown in Alg. 2.
As shown in Fig. 3, the graph $G = (V,E)$ includes all possible service placement decisions within $T$ time.
Each node represents the MEC node where the service is placed at time slot $t$ and we use node $S$ to represent user service's initial location. Each edge represents one possible service placement decision, and the weight on each edge is the perceived service cost (including computing delay, communication delay, and migration cost if $i\neq j$). A dummy node $D$ is used in our graph which helps to find a single shortest path. Obviously, the costs of edges connected to node $D$ are zero.

With the prediction of user's near-future information in $T$ time slots, the user's mobility over the time horizon can be given. By selecting different MEC nodes on user's moving trajectory, the cumulative service cost of the user in the time period can be computed. To illustrate that, the total weight of a path (e.g., the red line in Fig. 3) is user's accumulated service cost over the time horizon by selecting MEC nodes $S,c,b,b,d$ separately at each time slot.
By taking the shortest path from the source node $S$ to destination node $D$, we can easily find the optimal service placement solution to problem (20). The algorithm is described explicitly in Algorithm 2, in which the shortest-path (i.e., the optimal placement policy) for each frame can be found by using Dijkstra algorithm (i.e., line 17).
Since our problem has the optimal substructure property, we can iteratively use the optimal service placement strategy at current time slot to find the shortest path in the next time slot.
After iterations in time $T$, we select the minimum cumulative cost (i.e., line 22) to get the optimal service placement policy.
Searching for the minimum cumulative cost as the current optimal service placement strategy, the system needs to enumerate the possible configurations at most $O(N^2)$. Therefore, in $T$-time period, the time-complexity of Algorithm 3 is $O(N^2T)$.
\begin{figure}[tbp]
	\centering
	\includegraphics[width=3.5in,height=5.0in,clip,keepaspectratio]{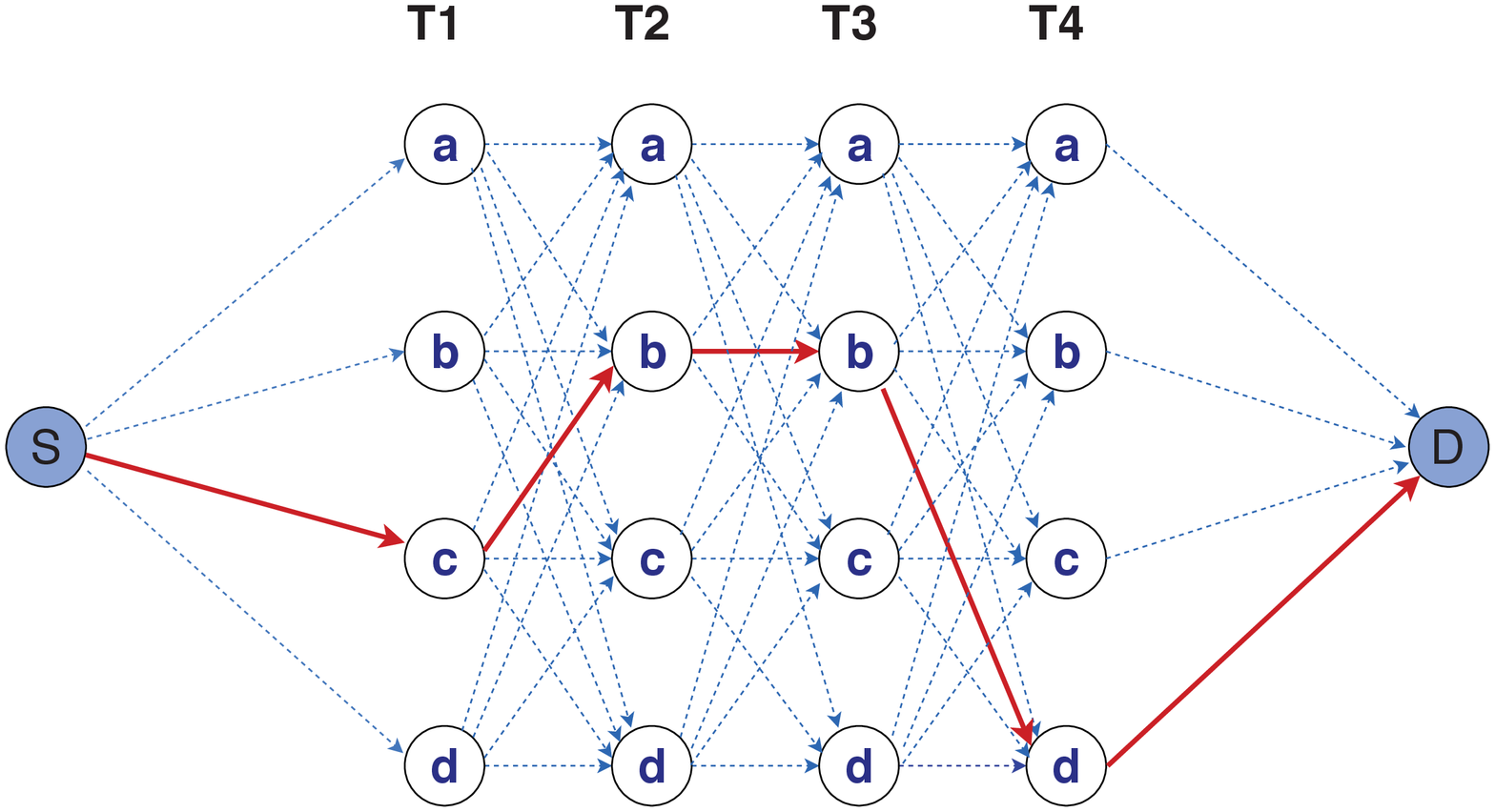}
	\caption{ The shortest-path problem transformation of $T-slot$ predictive service placement problem over $T=4$ time slot with 4 MEC nodes named a,b,c,d on the graph.}
	\label{fig:label}
\end{figure}

\begin{theorem} For any non-negative control parameter $V$, the long-term user-perceived latency and average queue length by proposed \textbf{PSP} algorithm satisfies:
	\begin{equation}
	L_{av}^{PSP} \triangleq \lim_{\widetilde{T}\to\infty}\sup\cfrac{1}{\widetilde{T}}\sum_{t=0}^{\widetilde{T}-1}\mathbb E{[L(t)]} \leq C(\theta)+\cfrac{B}{V},
	\end{equation}	
	\begin{equation}
	Q_{av}^{PSP} \triangleq \lim_{\widetilde{T}\to\infty}\sup\cfrac{1}{\widetilde{T}}\sum_{t=0}^{\widetilde{T}-1}\mathbb E\{Q(t)\}  \leq \cfrac{B+VC(\theta)}{\theta},
	\end{equation}
where $B$ is a constant defined in Lemma 1 and $\theta \in (0,\eta]$ where $\eta$ is defined in (12).
\end{theorem}

\subsection{Performance analysis of \textbf{PSP}}
Similar to \textbf{OSP}, we characterize the performance of \textbf{PSP} under the assumption that the user perceived latency $H_i(t)$ is independent and identically distributed (i.i.d.) over time slots and the condition in (12) is satisfied.

First, we define  $C(\theta)$ as the optimal expected time average user-perceived latency of problem (5), and $\theta>0$ is a finite constant that represents the distance between the time-averaged migration cost by some control policy and long-term cost budget.
Apparently, we have $\lim_{ \theta \to 0 }C(\theta)=L_{av}^{*}$.

Let $L_{av}^{PSP}$ and $Q_{av}^{PSP}$ be the long-term average user-perceived latency and average queue backlog in \textbf{PSP}, respectively. We state the following theorem (the detailed proof is given in our online technical report \cite{Appendix}).

Note that \textbf{PSP} achieves similar performance bounds as \textbf{OSP}. In particular, (a) when $\theta$ approaches 0, the bound for the perceived latency achieved by \textbf{PSP} is equal to that of \textbf{OSP} in (13). That is,

\begin{equation}
\lim_{ \theta \to 0 }(C(\theta)+\cfrac{B}{V}) \leq L_{av}^{*} +\cfrac{B}{V}.
\end{equation}
(b) When $\theta = \eta$, the average queue length of \textbf{PSP} satisfies:
\begin{equation}
Q_{av}^{PSP} \leq \cfrac{B+VC(\eta)}{\eta} \leq \cfrac{B+VL_{\max}}{\eta}.
\end{equation}
where $L_{\max}=\max_t L(t)$, and the right bound is the same as the one specified in (14).

Due to the the fact that \textbf{PSP} operates in a larger timescale, and decides the service placement policy in a joint consideration over the whole frame, the actual gaps between the two sides of the inequalities (23) and (24) are usually much larger than those in (13) and (14). That is to say, \textbf{PSP} can achieve a better cost-delay trade-off than \textbf{OSP}. In section VII, we will illustrate the performance of our proposed algorithm through extensive performance evaluation, which shows an up-to 50.4$\%$ user-perceived latency reduction at the same migration cost budget over existing benchmarks without predictions. 
\section{Extension with Historical Queue Information}
In previous sections, we consider the predictive service placement (\textbf{PSP}) algorithm design using the predicted information. As a further enhancement, we propose a weight-update method for \textbf{PSP} named \textbf{PSP-WU} to exploit the historical queue backlog information.

\subsection{The Weight-update Method}
In \textbf{PSP}, we design the algorithm only based on the current and predicted future information, without the information for previous frames. Specifically, we use the value of $Q(kT)$ to present the queue-length  at each time slot in frame $k$. 
Rather than using only the queue information at the current frame and being memoryless of the past frames' information, we propose a new weight adjustment scheme for the queue in each frame named \textbf{PSP-WU} to exploit the extensive historical queue information.
 
Different from the previous queue-length update method, we introduce a new weight $W(t)$.
First we update the queue-length following (6). Let $\Delta_Q \triangleq Q(t+1)-Q(t)$ be the resultant queue-length change. Next, we update the weights in the following:
\begin{equation}
W(t+1)=W(t)+\Delta_Q+\beta[W(t)-W(t-1)]^+.
\end{equation}
The factor $\beta\in[0,1]$ is the parameter to set the weight of historical information. When $\beta=0$, it reduces to the original queue-length update method. Thus, when $\beta>0$, we can incorporate the queue backlog information of the previous slot in a recursive and cumulative way. 
\begin{algorithm}[tp]
	\caption{ Weight-update based predictive service placement}
	\begin{small}
		\begin{algorithmic}[1]
			\STATE 	Define variables $i$ and $j$ to represent the possible decision respectively in the current and previous time slot.	
			\STATE Define array $\pi_{i}$ and $\zeta_{j}$ for all $i$,$j$, where $\pi_{i}$ and $\zeta_{j}$ are the optimal decisions from the beginning $t$ to the current and previous time slot when the decision is $i$ respectively.
			\STATE Define variables $\phi_{i}$ and $\varphi_{i}$ for all $i$ to count the cumulative cost from the beginning $t$ respectively to the current and previous time slot when the decision  is $i$.
			\STATE Define variables $D(i,t)$ to represent the service cost when the decision is $i$ at time slot $t$.
			\STATE  Initialize the cost queue backlog $Q(0) = 0$. Choose $\beta\in[0,1]$.  Set $W(0)=W(-1)=0$.
			\FOR {each frame  $k \in \{0,1,2,...,K-1\} $}
			\FOR { all $i$}
			\STATE  $\pi_{i}=\emptyset$.
			\STATE  $\phi_{i} = 0$.
			\ENDFOR
			\FOR { $t = kT,kT+1,...,kT+T-1$}
			\FOR { all $i$} 
			\STATE $\zeta_{i}$ = $\pi_{i}$.
			\STATE $\varphi_{i}$ = $\phi_{i}$.
			\ENDFOR
			\FOR {all $i$} 
			\STATE $j_{opt}$ = $\arg\min_{j}\{\varphi_{j}+D(i,t)\}$.
			\STATE $\pi_{i}(kT,...,t)=\zeta_{j_{opt}}\cup i$.
			\STATE $\phi_{i}=\varphi_{j_{opt}}+D(i,t)$.
			\ENDFOR
			\ENDFOR
			\STATE  $i_{opt}=\arg\min_{i}\phi_{i}$.
			\STATE  The optimal policy in each frame k is: $\pi_{i_{opt}}( kT,kT+1,...,kT+T-1)$.
			\STATE $\mathbf{Update}$ the virtual queue: run (6) based on $\mathbf{x^{*}(\tau)}$ at each time slot $\tau$ in the frame $k$.
			\STATE $\mathbf{Update}$ the weight: run (25) based on the virtual queue and $\mathbf{x^{*}(\tau)}$ at each time slot $\tau$ in the frame $k$.
			\ENDFOR
		\end{algorithmic}
	\end{small}
\end{algorithm}
\subsection{Weight-update based Predictive Service Placement}
Next, we present the weight-update based $T$-slot predictive service placement (\textbf{PSP-WU}) algorithm by solving the following optimization problem in each time frame $k \in \{0,1,2,...,K-1\} $ by replacing the queue parameters to the weight parameters in the \textbf{PSP} problem in (20) as follows:
\begin{equation}
\begin{aligned}
\min \ \sum_{ \tau=kT}^{kT+T-1}& \{W(kT)(E(\tau)-E_{avg}+\theta[(k+1)T-\tau] )\\&+VL(\tau)\}  \\ 
&\text{s.t.} \quad (1)-(3),(6),(25).
\end{aligned}     
\end{equation}

The \textbf{PSP-WU} algorithm is summarized in Alg. 3. Note that it is very challenging to theoretically derive the performance bound for \textbf{PSP-WU} algorithm due to the accumulated and coupling natures in weight update therein. We will consider it in a future work. Numerical results demonstrate that \textbf{PSP-WU} is effective and helps to reduce the queue length while improving the quality of user-perceived latency.

\section{PERFORMANCE EVALUATION}
In this section, we evaluate the performance of online service placement (\textbf{OSP}) algorithm, $T$-slot predictive service placement (\textbf{PSP}) algorithm and weight-update $T$-slot predictive service placement (\textbf{PSP-WU}) algorithm using realistic user mobility trace to verify the theoretical results and gain useful insights. 

In our study, we use a real-world trace of mobile users using Twidere (an open-source Android Twitter)\footnote{https://github.com/TwidereProject/Twidere-Android}. We average our experiments with 100 users who have consecutive locations records. The dataset contains information about user's locations obtained from the GPS timestamp, which are highly dispersed. We select the users who moved from 30.8 to 31.4 longitude and 121.2 to 121.6 latitude, and record the timestamp every hour from 7 a.m. to 8 p.m for 100 days. 
We use K-Means clustering algorithm to automatically cluster the user's positions points into multiple regions (i.e., regular locations). By analyzing these timestamps and user's positions, we find that the user's positions every hour of the day are regular. So we cluster the user's locations into 6 different regions. 
Every region has one base station to which the MEC node will be attached. Table II lists the details of simulation set up about service
requirement and migrtaion cost state.
\begin{table}[tp] 
	\caption{Simulation  Set up about Service
		Requirement and Migrtaion Cost}
	\centering	
		\normalsize
	\begin{tabular}{ll} 
		\toprule 
		Description & Value Range\\ 
		\midrule 
		Task input size & [5, 10]MB \\ 
		Computing workload& [2, 20]G CPU Cycles\\
		Cellular bandwidth & [5, 10]MHz \\ 
		Computing capacity of MEC node & [5, 10]GHZ\\ 
		Data size of the service container  & [25, 50]MB \\ 
		Service migration unit cost&  [2, 10]Dollars/GB\\ 
		\bottomrule 
	\end{tabular}
\end{table}

For location prediction, we utilize the popular deep learning method of long short-term memory (LSTM) \cite{l27} to obtain the results. 
The hidden unit size of the LSTM is 128, for which we can generate the best prediction performance.
We implement our experiments on datasets, which are splited into the training set (60$\%$), the development set and the test set (40$\%$).
In the experiments, we use Adam optimization method to minimize the loss of mean square error on the training set.
Meanwhile, we use mean square error as the metric to evaluate our model. 
During the training process, we train the model for fixed epochs and monitor its performance on the validation set. 
Once the training is finished, we will select the model with the best mse score on the validation set as our final model, which can be saved in the training process, and evaluate its performance on the test set.
When $W=1$ (i.e., the prediction window size is 1 and time frame size $T=2$), our prediction accuracy is 90.3$\%$, and prediction accuracy are 83.9$\%$ and 54.8$\%$ when $W=2$ and $W=3$, respectively.  We also use another two methods. One is autoregressive integrated moving average model (ARIMA) \cite{ARIMA} and the other is simple moving average (SMA) \cite{SMA}. Table III lists the prediction accuracy with different prediction window sizes. LSTM achieves the best performance than the other two comparing methods. In our experiment, we use the prediction result of LSTM to evaluate the performance.
\begin{table}[tp] 
	\caption{Location prediction accuracy with different prediction window sizes}
	\centering	
	\normalsize
	\begin{tabular}{llll} 
		\toprule 
		  Methods	& $W=1$ &  $W=2$ &  $W=3$  \\
		\midrule 
		LSTM & 90.4$\%$  & 83.9$\%$  &54.8$\%$ \\
		ARIMA & 88.5$\%$ & 80.8$\%$  & 50.9$\%$\\
		SMA & 35.5$\%$ & 10.2$\%$  & 0.2$\%$\\
		\bottomrule 
	\end{tabular}
\end{table}
\subsection{Performance Benchmark}
In order to evaluate our proposed algorithm, we compare it with two representative situations and two greedy approaches. Furthermore, the details of the 4 benchmarks are outlined in the following:
\begin{enumerate}
	\item Always Migration Algorithm (\textbf{AM}): no matter what the distribution of mobile user is, the service is always migrated to execute on its nearest MEC node.
	\item No Migration Algorithm (\textbf{NM}): keep the initial assignment policy unchanged, no matter where the user is.
	\item Lazy Migration Algorithm (\textbf{LM}): the basic idea of LM algorithm is to postpone service migration until the cumulative non-migration latency has significantly exceeded the potential migration cost. In this way, the service will not be migrated frequently. 
	The LM is widely used in literature such as \cite{lazy}.
	\item Predictive Lazy Migration Algorithm (\textbf{PLM}): 
	inspired by \cite{Plazy}, we use PLM algorithm as a benchmark. 
	For each time slot, if service is not placed at user's nearest MEC node, then we will make trade-off between the service migration cost and the possible service latency of the next time slot in non-migration case. By leveraging the predicted location in the next time slot, we can make wiser service migration decision in current time slot. 
\end{enumerate}

\subsection{Performance Analysis}
There is no doubt that the key challenge of the long-term dynamic service placement problem is to optimize the user-perceived latency and migration trade-off in a cost-efficient manner, which guides the following analysis for our proposed algorithms: reactive service placement \textbf{OSP} algorithm, predictive service placement (\textbf{PSP}) algorithm, and weight-update predictive service placement (\textbf{PSP-WU}) algorithm.

\emph{Average user-perceived latency optimality}. In order to analyse key elements affecting user-perceived latency, we set the long-term time-averaged migration cost budget for the network operator as 417 cost units, the number of MEC nodes are 6, and the control parameter $\theta$ and $\beta$ are set as 50 and 0.65 respectively.  

\begin{figure*}[tp]
	\begin{minipage}[t]{0.32\linewidth}
		\centering
		\includegraphics[height=2.25in,width=2.35in]{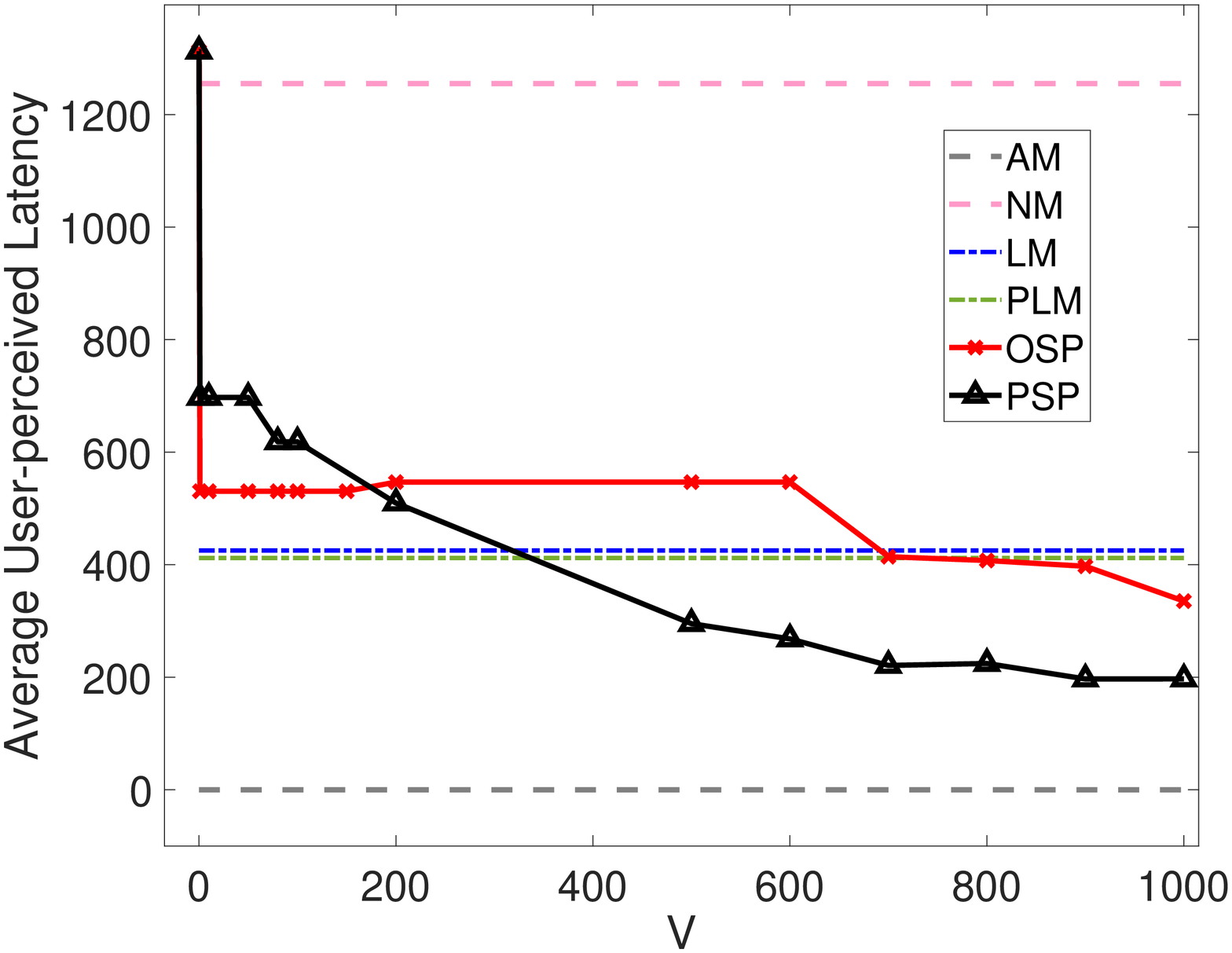}
		\caption{Average latency with different values of control parameter $V$}
	\end{minipage}
	\hfill
	\begin{minipage}[t]{0.32\linewidth}
		\centering
		\includegraphics[height=2.25in,width=2.35in]{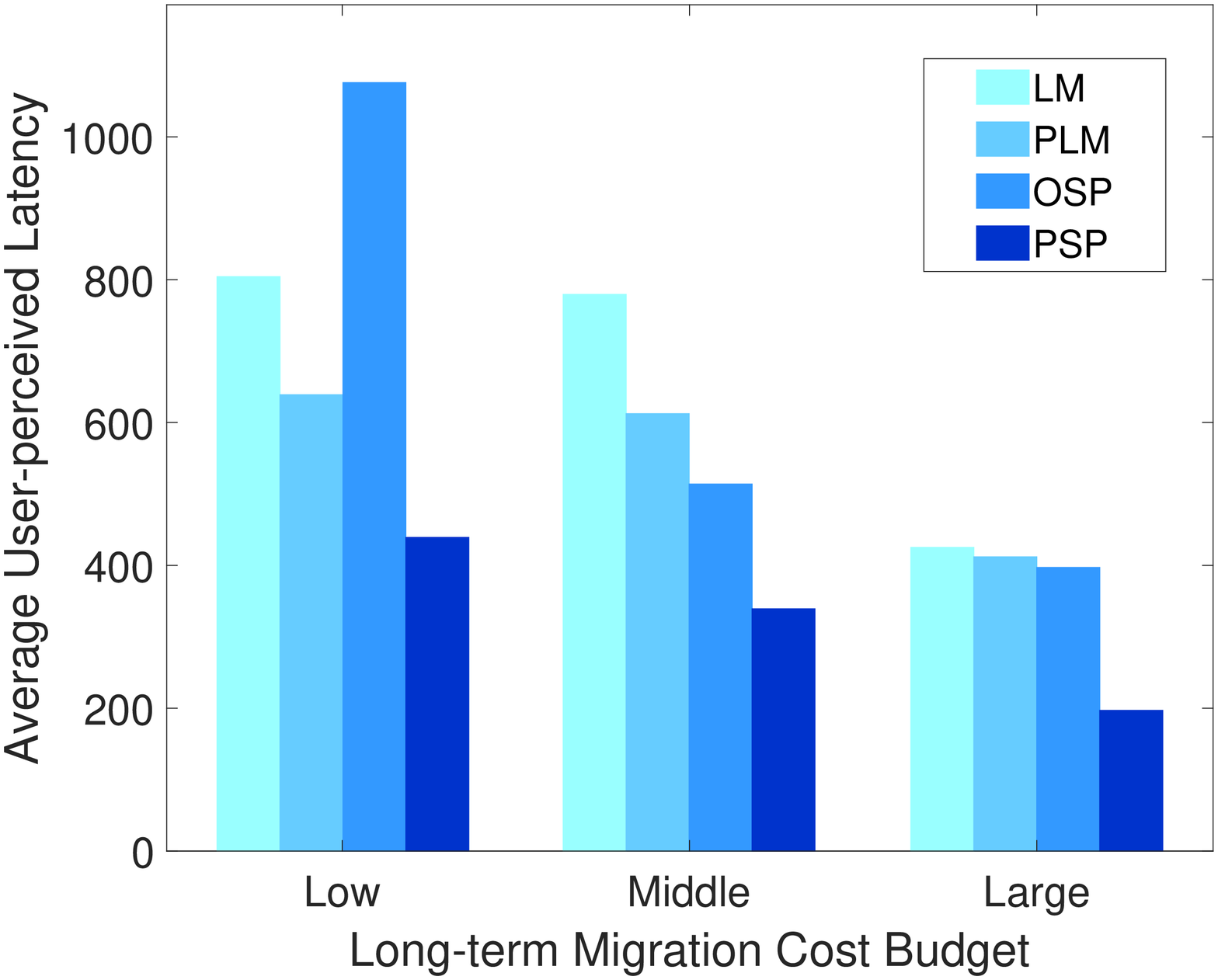}
		\caption{Average latency with different long-term budgets $E_{avg}$}
	\end{minipage}
	\hfill
	\begin{minipage}[t]{0.33\linewidth}
		\centering
		\includegraphics[height=2.35in,width=2.35in]{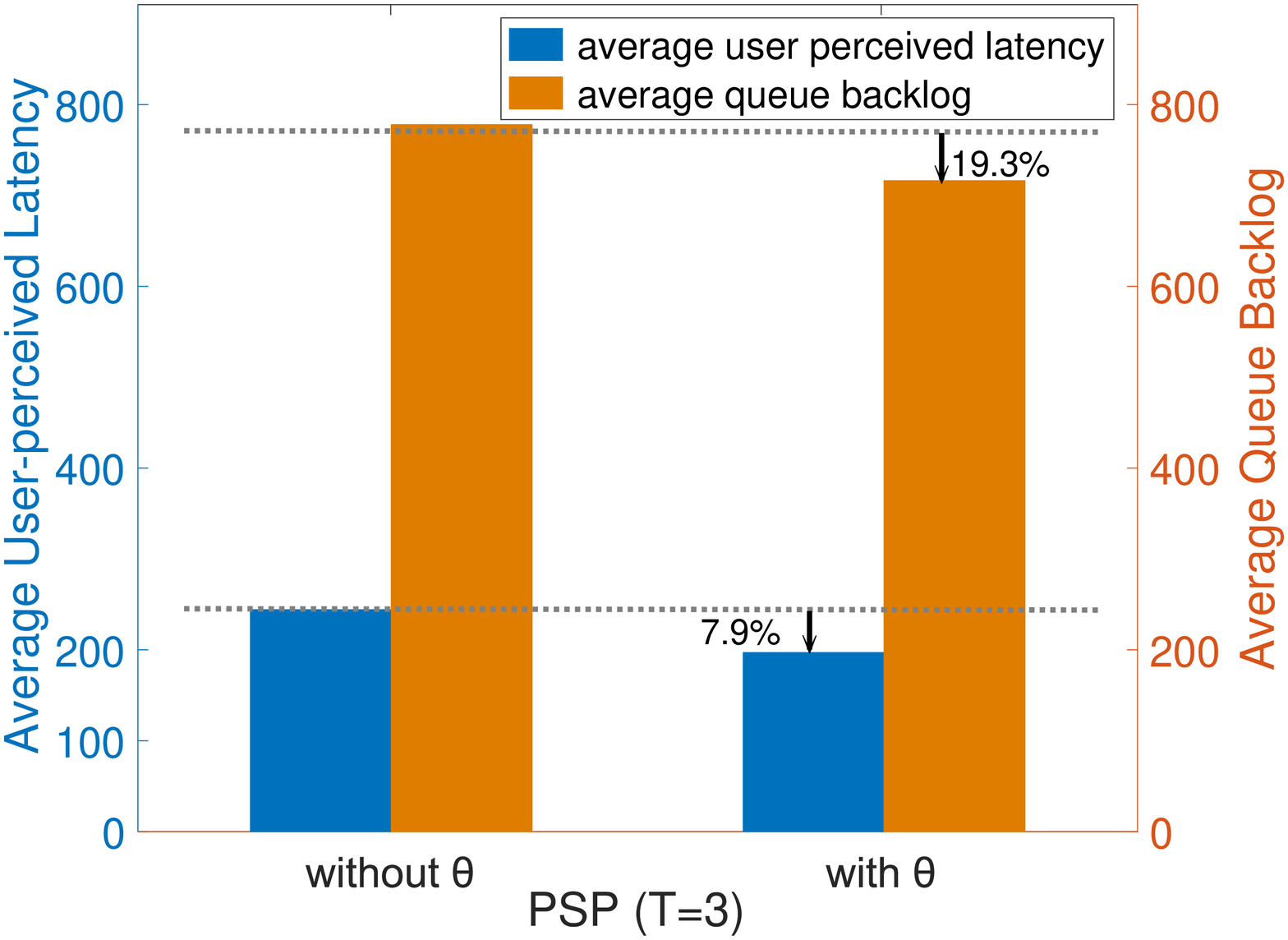}
		\caption{\textbf{PSP}'s performance in $\theta$}
	\end{minipage}
\end{figure*}
\begin{figure*}[tp]
	\begin{minipage}[t]{0.32\linewidth}
		\centering
		\includegraphics[height=2.25in,width=2.35in]{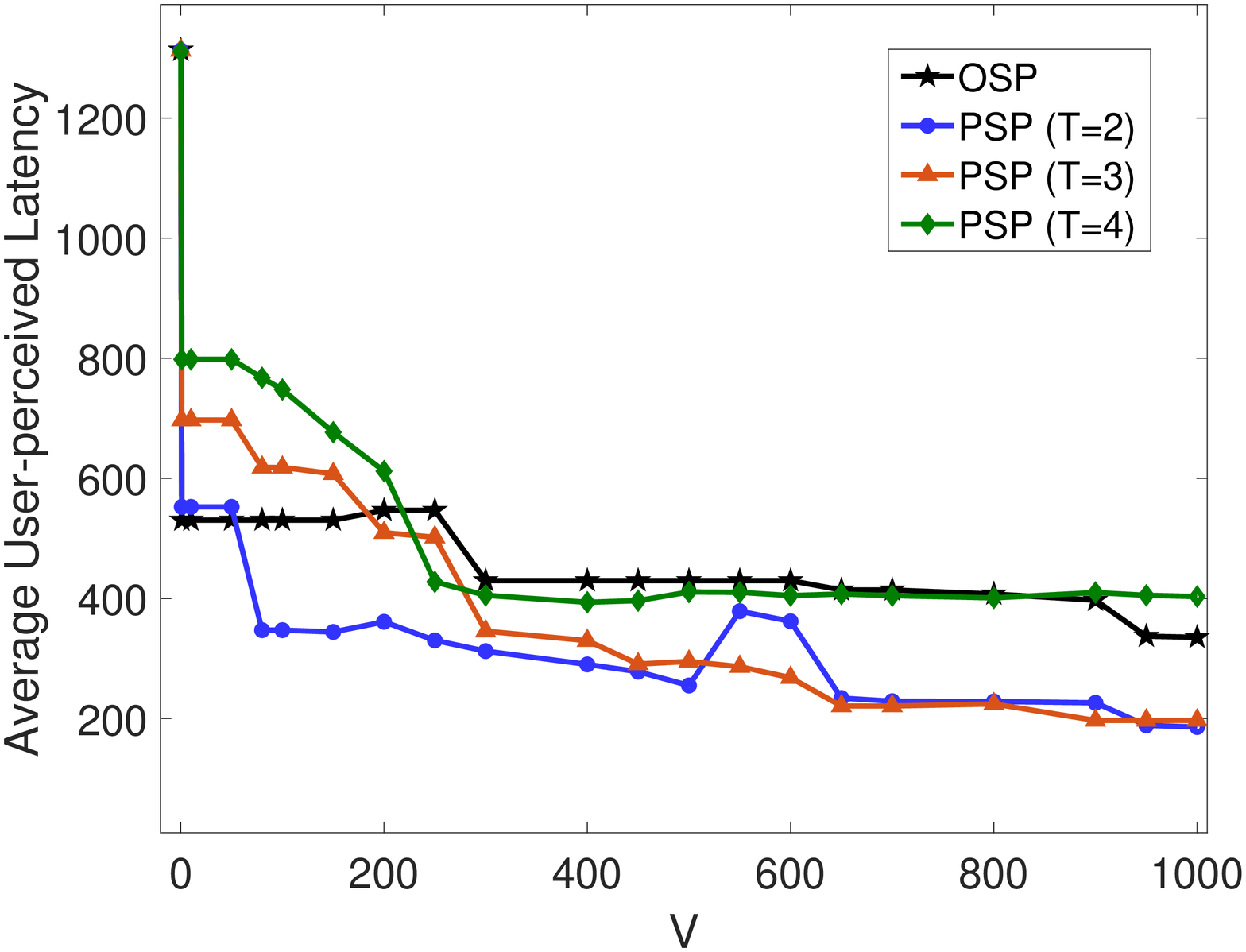}
		\caption{Average latency with different values of control parameter $V$}
	\end{minipage}
	\hfill
	\begin{minipage}[t]{0.32\linewidth}
		\centering
		\includegraphics[height=2.25in,width=2.35in]{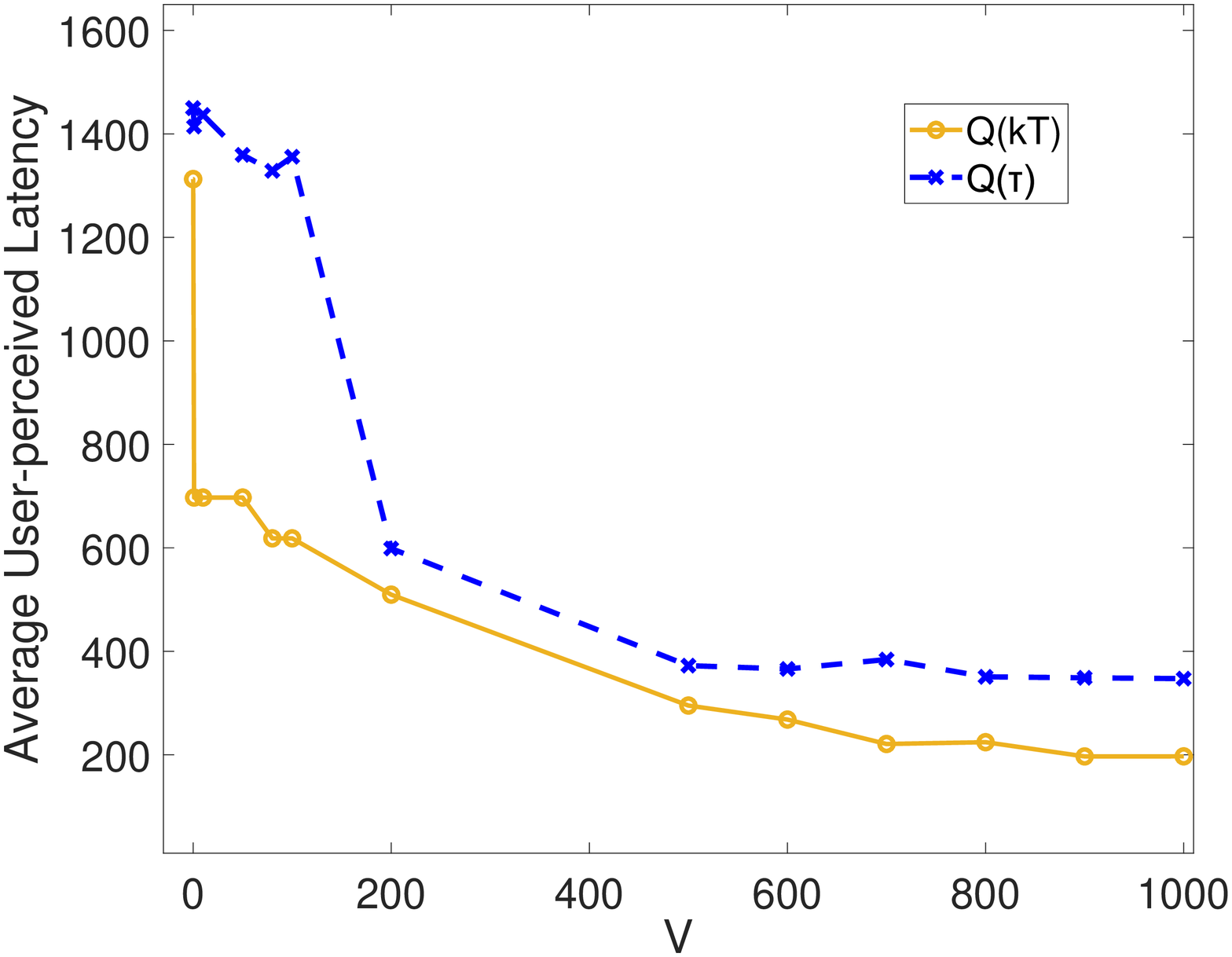}
		\caption{Average latency performance under $Q(\tau)$ and $Q(kT)$}
	\end{minipage}
	\hfill
	\begin{minipage}[t]{0.33\linewidth}
		\centering
		\includegraphics[height=2.25in,width=2.35in]{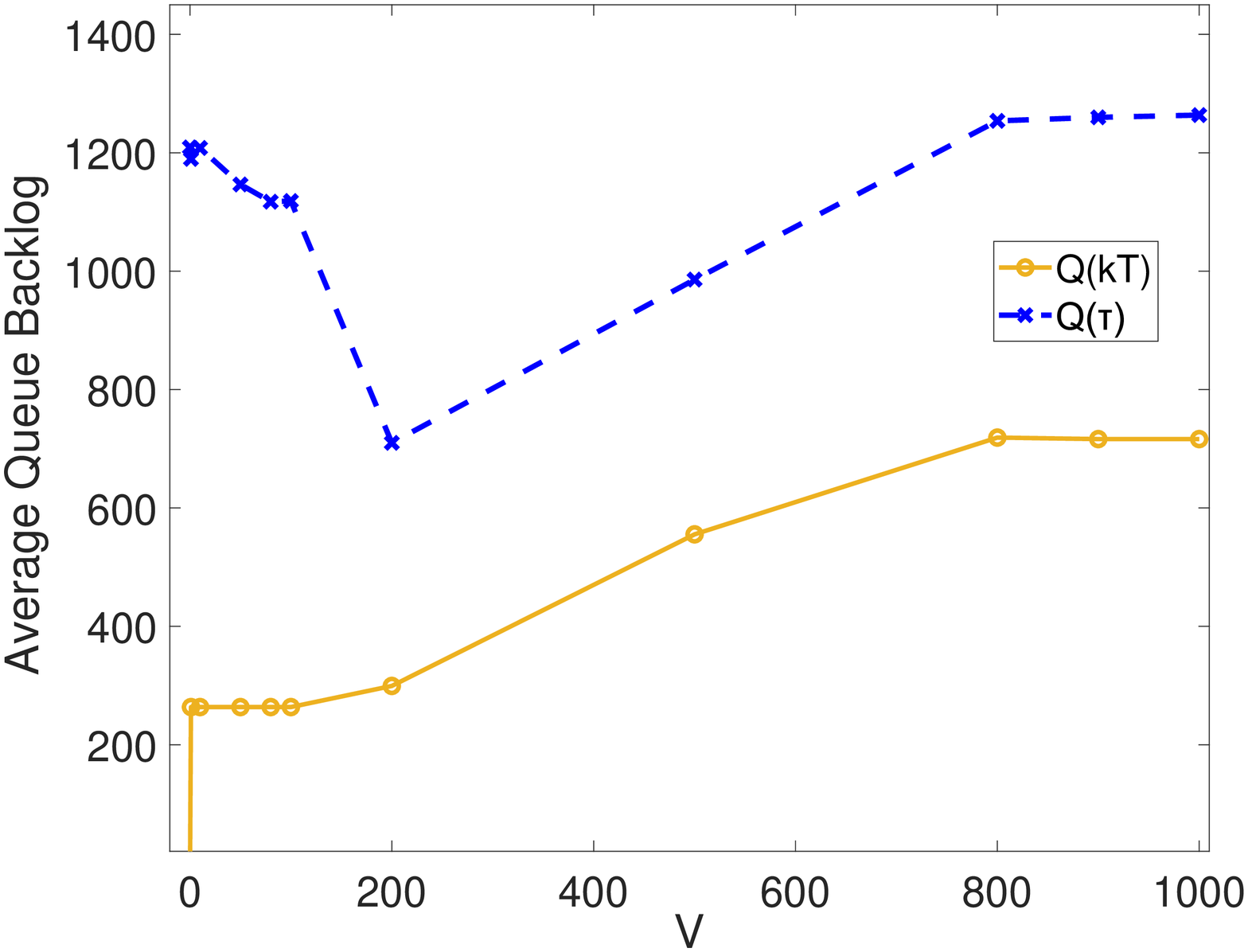}
		\caption{Average Queue Backlog under $Q(\tau)$ and $Q(kT)$}
	\end{minipage}
\end{figure*}
Fig. 4 plots the average latency with different values of control parameter $V$ under various online algorithms. We can observe that the average latency decreases with $V$ increasing. When $V$ is large enough, \textbf{PSP}'s latency will approach a minimum value and \textbf{OSP}'s latency will keep reducing until it gets to a minimum. This is because the larger the $V$ is, the more important the user-perceived latency becomes in the optimization. In other words, as $V$ increases, the service should be placed as close as possible to guarantee the latency performance. 
Except for several fluctuations, this experimental results are in general agreement with the theoretical analysis in Theorem 1 and  Theorem 2 that the time-averaged latency performance is proportional to $1/V$. 
And we can see the performance of PLM is a little better than LM. NM gets the worst performance because in NM no matter where the user is, the service is always placed at the same MEC node, such that user's latency performance is poor. AM gets the best performance because the service is always placed at the nearest MEC node to serve user.
Compared with these benchmarks, our algorithms do have remarkable improvements in average latency performance. When $V=900$, \textbf{PSP}'s improvement is 50.4$\%$. 

Intuitively speaking, a larger migration cost budget can supply further efforts to the optimization of service placement. As illustrated in Fig. 5, we set low as 167 cost units, middle as 260 cost units, and large as 417 cost units. We can see with the cost budget increasing, \textbf{PSP} has more notable improvements compared with LM, PLM and OSP. For example, given a budget of 260 cost units, the latency reduction ratio of \textbf{PSP} is 30.4$\%$. 

\emph{Performance in $\theta$}. Fig. 6 suggests that $\theta$ can capture the variance of queue length within each frame. By using $\theta$, the latency reduction ratio is 7.9$\%$ while the average queue backlog reduction ratio is 19.3$\%$, both of them verify the intuition in Section V.
\begin{figure*}[tbp]
	\begin{minipage}[t]{0.32\linewidth}
		\centering
		\includegraphics[height=2.25in,width=2.35in]{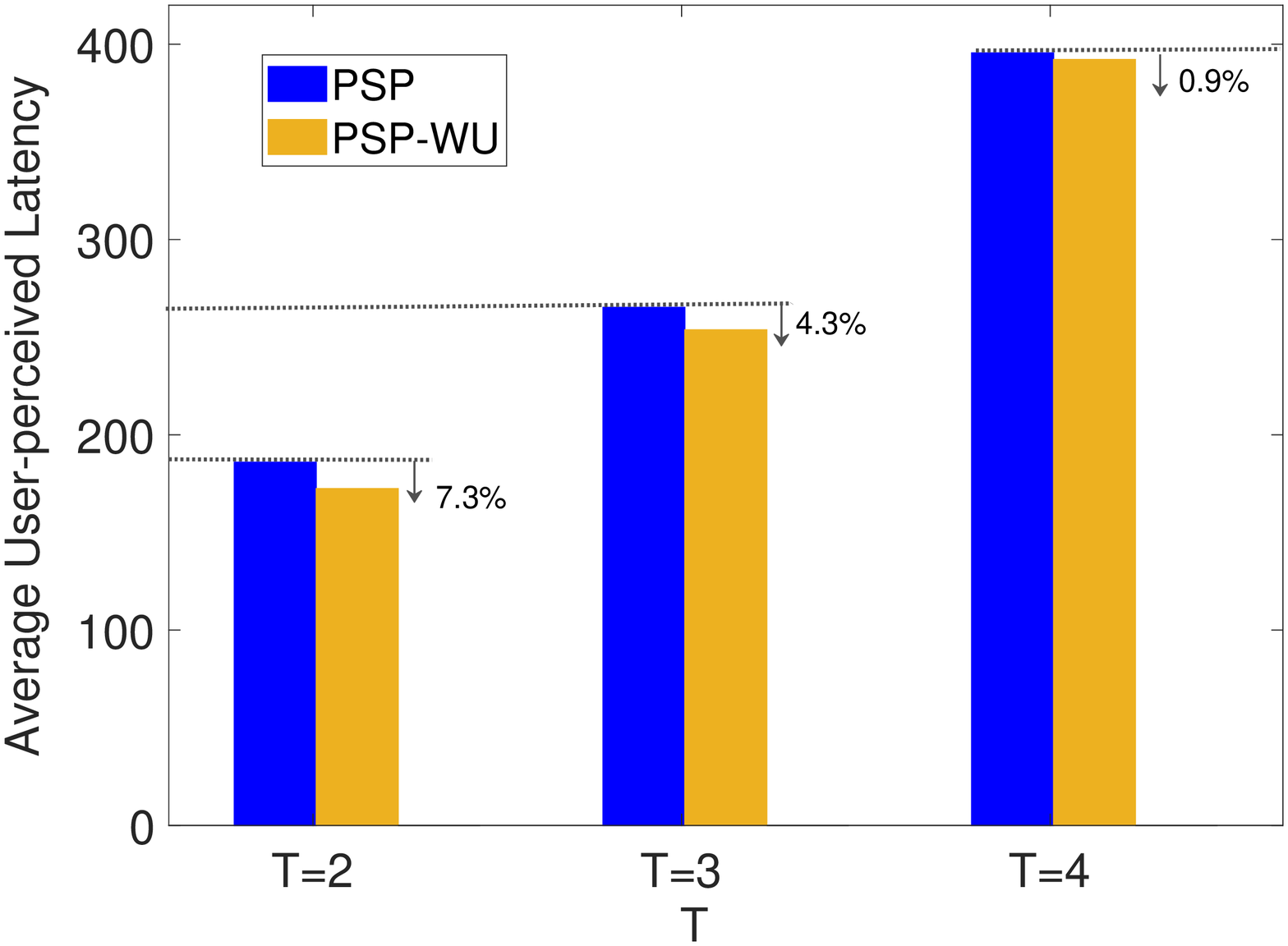}
		\caption{Average latency in PSP and PSP-WU }
	\end{minipage}
	\hfill
	\begin{minipage}[t]{0.33\linewidth}
		\centering
		\includegraphics[height=2.25in,width=2.35in]{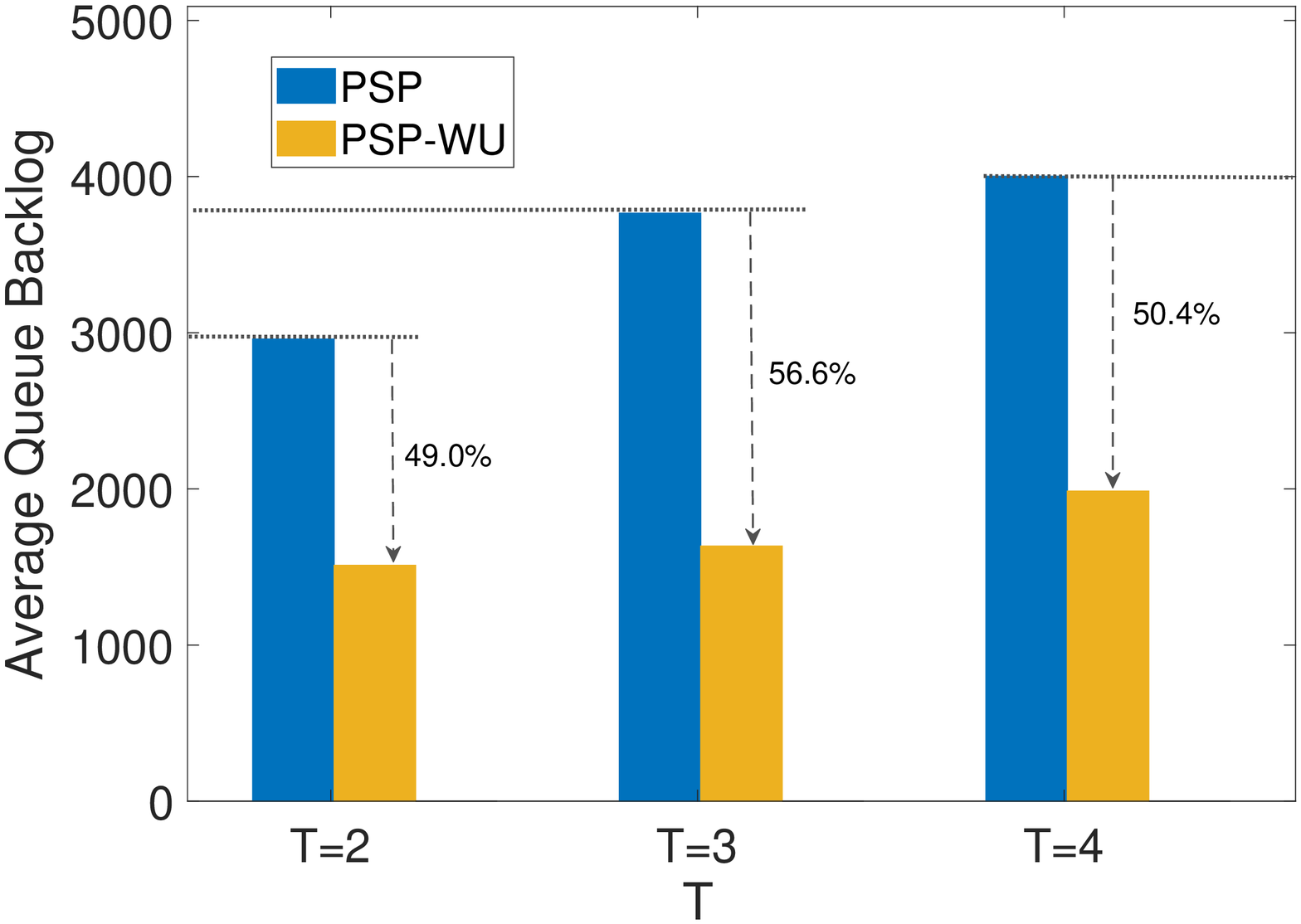}
		\caption{Average queue backlog in PSP and PSP-WU}
	\end{minipage}
	\hfill
	\begin{minipage}[t]{0.32\linewidth}
		\centering
		\includegraphics[height=2.25in,width=2.35in]{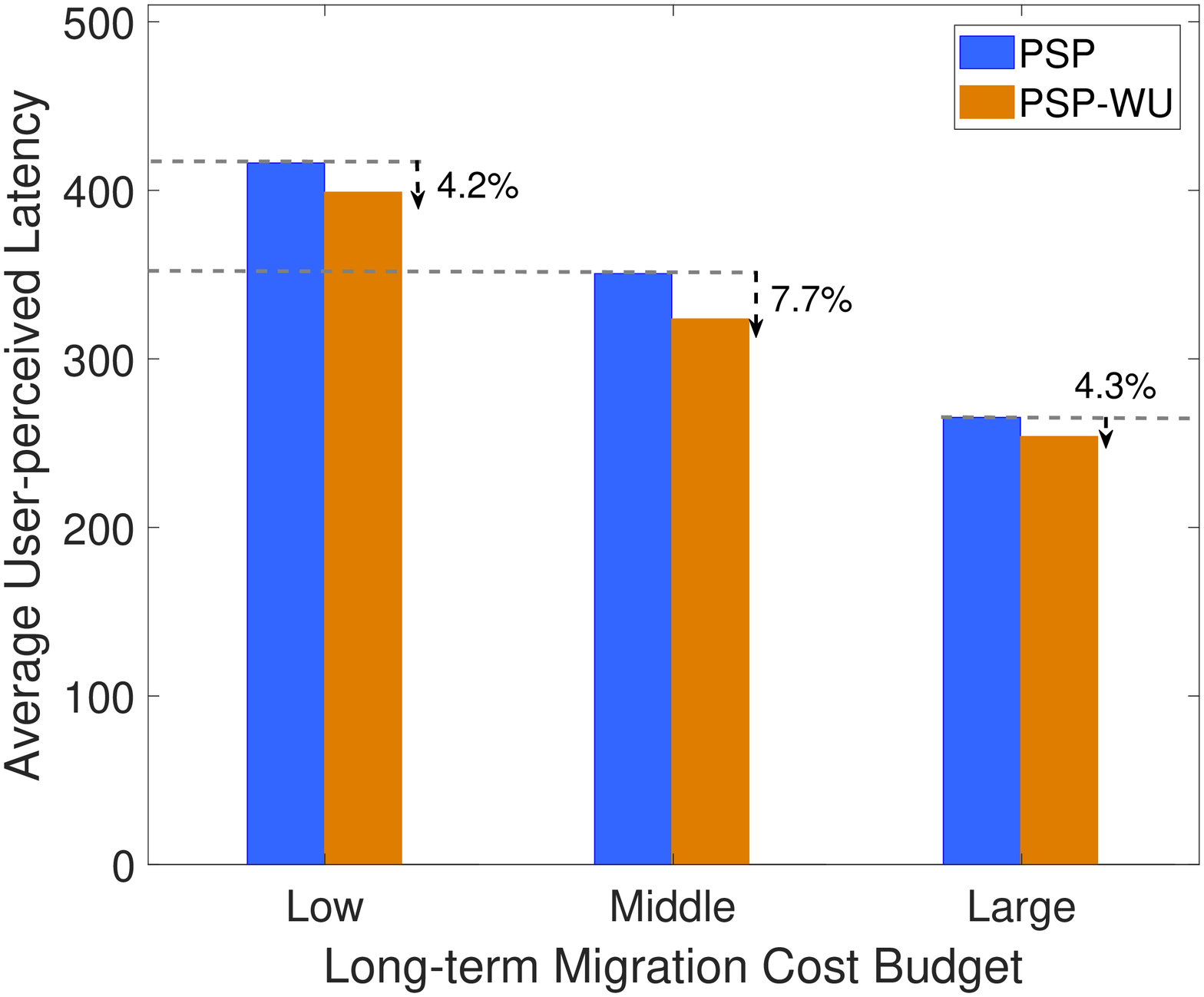}
		\caption{Average latency in PSP and PSP-WU under different $E_{avg}$}
	\end{minipage}
	\hfill
\end{figure*}
\begin{figure*}[tbp]
	\begin{minipage}[t]{0.32\linewidth}
		\centering
		\includegraphics[height=2.38in,width=2.35in]{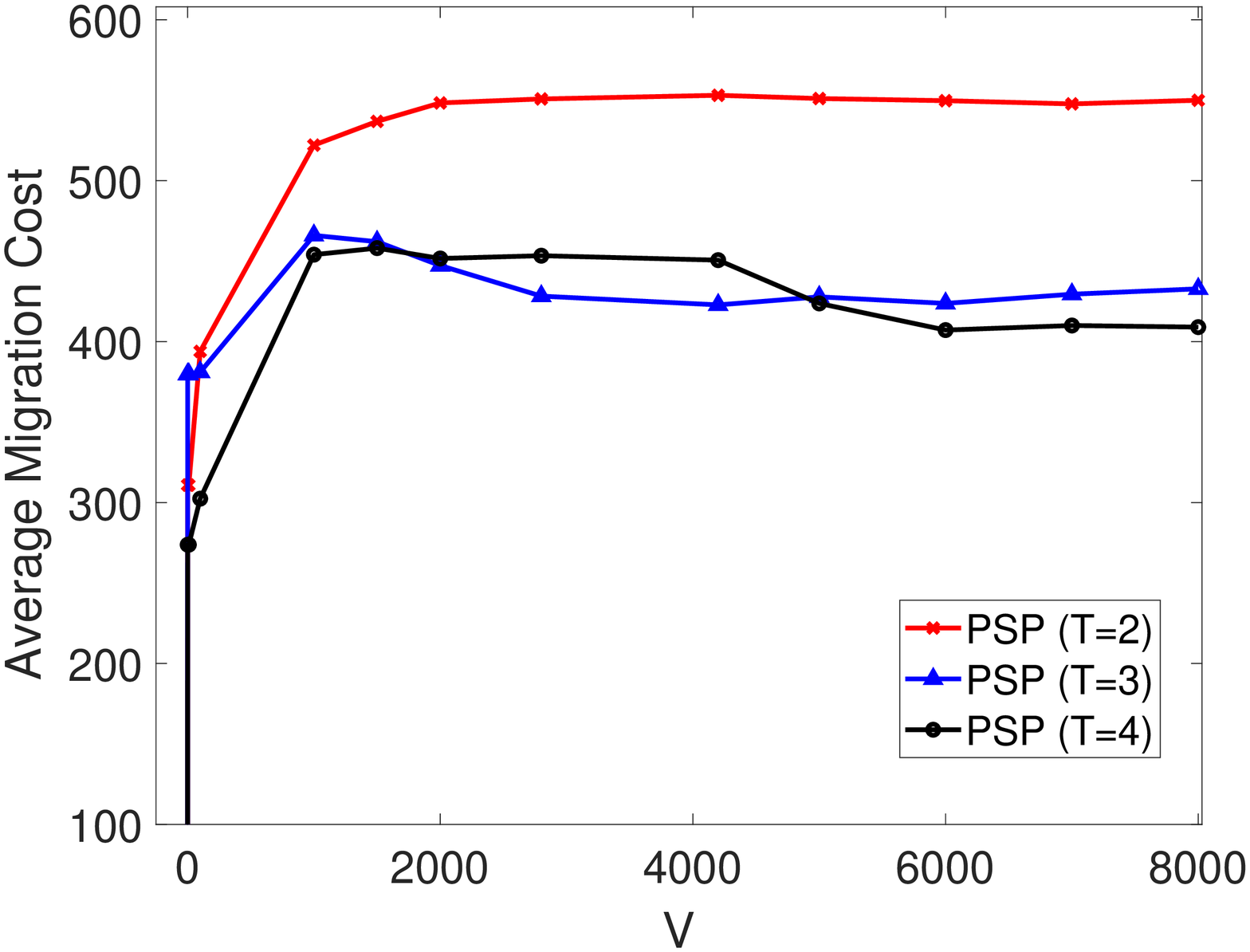}
		\caption{Average migration cost with different values of  $V$}
	\end{minipage}
	\hfill
	\begin{minipage}[t]{0.33\linewidth}
		\centering
		\includegraphics[height=2.25in,width=2.35in]{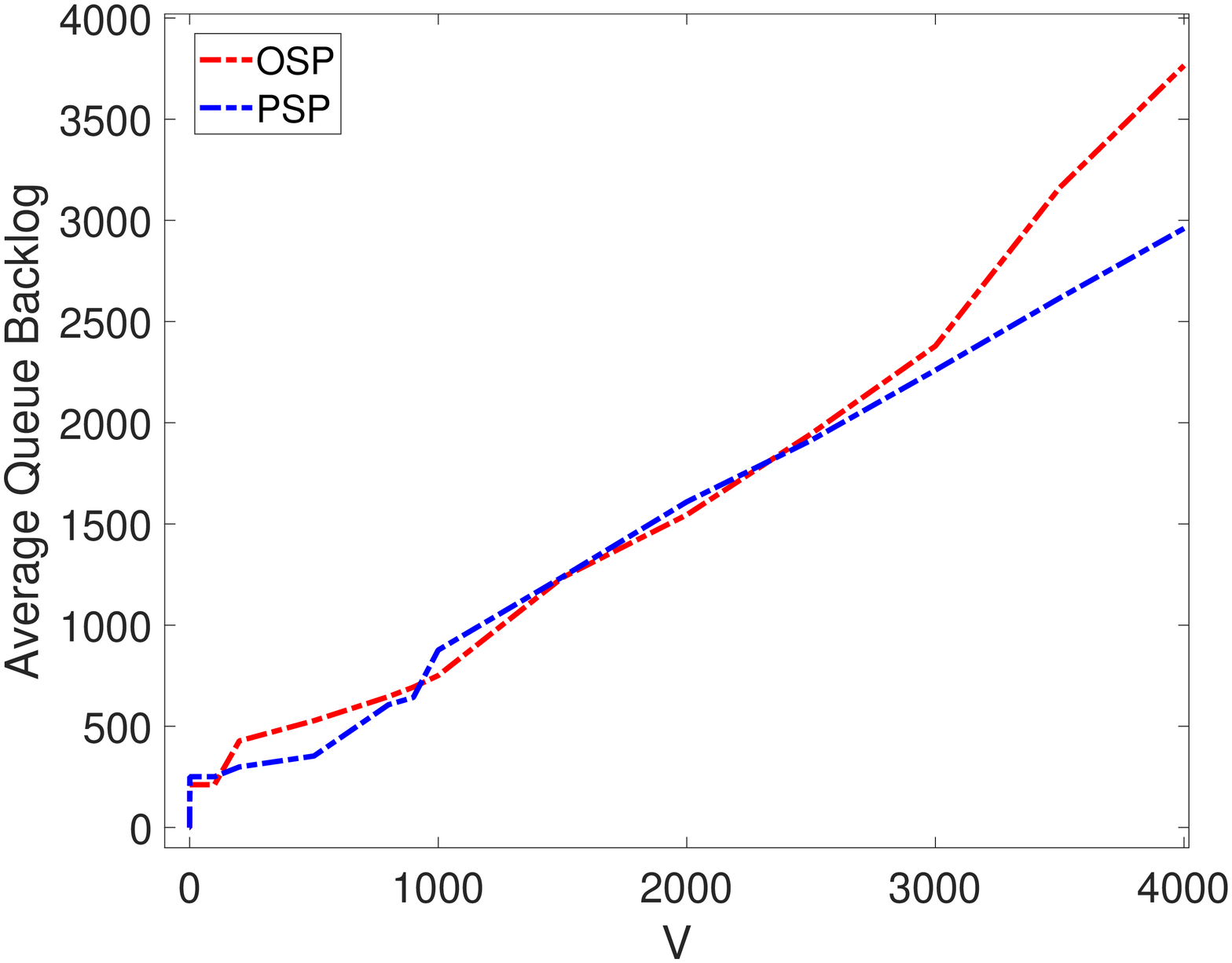}
		\caption{Average migration cost queue with different values of $V$ }
	\end{minipage}
	\hfill
	\begin{minipage}[t]{0.32\linewidth}
		\centering
		\includegraphics[height=2.25in,width=2.4in]{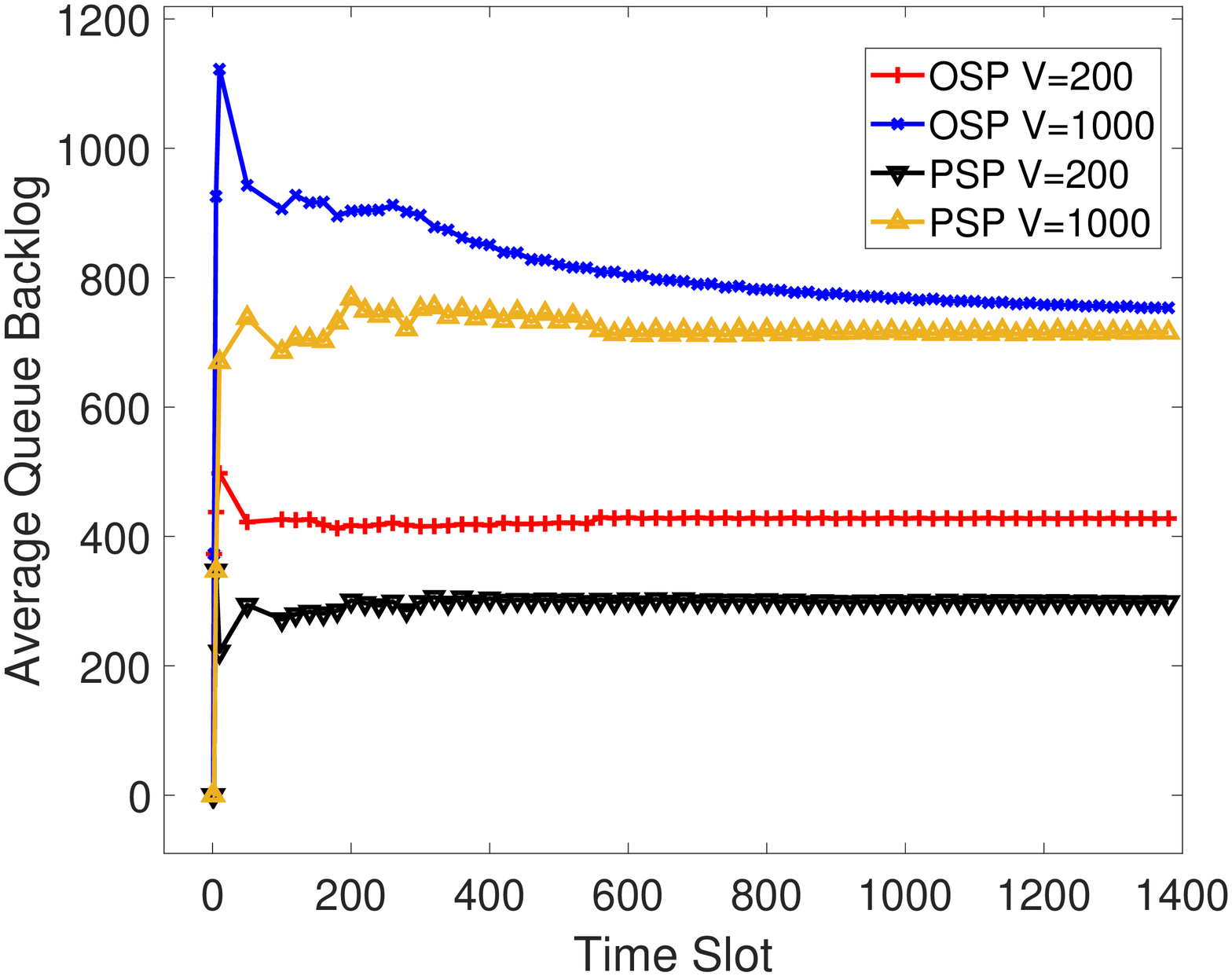}
		\caption{Average migration cost  with different values of $V$ at different time slot}
	\end{minipage}
	\hfill
\end{figure*}
\emph{Performance in different $T$}. Fig. 7 shows the average user-perceived latency with different values of control parameter $V$ under various $T$. We can observe that the average latency decreases with $V$ increasing, and gradually approaches to a minimum value in all $T$. 
Comparing average latency performance in different period $T$, we see $T=4$ (the prediction window size $W=3$) gains the worst performance, this is due to $T=4$'s low accuracy of prediction, resulting in a poor performance. While $T=3$ (the prediction window size $W=2$) and $T=2$ (the prediction window size $W=1$) gain the similar performance, and are  both better than the performance of \textbf{OSP}. This shows that when the prediction errors are modest, the performance achieved by \textbf{PSP} can be  acceptable and better than the performance of \textbf{OSP}. 
Here we acknowledge that since the available training sample is small and we adopt a simple all-connected output window position information to predict the user mobility, the potential of LSTM may has not been fully unlocked.

\emph{\textbf{PSP}'s performance in $Q(kT)$.}  As shown in Fig. 8 and Fig. 9, using $Q(kT)$ achieves better performance in both averaged user-perceived latency and queue backlog, which proves the  effectiveness of taking an approximation by setting the future queue backlogs as their current backlogs  to address the prediction errors and to reduce the complexity. 

\subsection{Performance of \textbf{PSP-WU} method}
To illustrate the performance in \textbf{PSP-WU} of user-perceived latency under different values of $T$  and the reduction ratio in average queue backlog, we depict the situation when $V=4000$ the average latency performance and average queue backlog with different values of $T$ and with different long-term budget values of $E_{avg}$. We set $E_{avg}$ as 417 cost units to compare the performance of \textbf{PSP-WU} with \textbf{PSP} under different values of $T$ and set $T=3$ to compare these two methods under different values of $E_{avg}$. 

As illustrated in Fig. 10,  we can see  \textbf{PSP-WU} always achieves smaller user-perceived latency compared to textbf{PS} under all the values of $T$.  For example, when $T$=2, the user-perceived latency reduction ratio of \textbf{PSP-WU} is 7.3$\%$.
From Fig. 11, we can see the perfect performance of \textbf{PSP-WU} method in average queue backlog. When $T=3$, the reduction ratio of \textbf{PSP-WU} is 56.8$\%$ which even decreases the queue backlog more than half of the average queue backlog in \textbf{PSP}. 
In Fig. 12, we can see with the cost budget increasing, \textbf{PSP-WU} has more notable improvements compared with \textbf{PSP}. For example, given a budget of 260 cost units, the latency reduction ratio of \textbf{PSP-WU} is 7.7$\%$. 

From Fig. 10, Fig. 11 and Fig. 12, we observe that \textbf{PSP-WU} achieves  smaller average user-perceived latency and queue backlog compared to \textbf{PSP}, thus we  see the efficiency of the \textbf{PSP-WU} method by leveraging the historical information.

\subsection{Cost trade-off}
\emph{Queue stability.} Fig. 13 plots the average migration cost with different values of $V$ under various $T$. The average migration cost in $T=4$ is smaller and more stable than that in $T=3$, while the average migration cost in $T=2$ is the biggest.
We can see  from Fig. 7 and Fig. 13 when $T=3$, the user-perceived latency decreases faster and the average migration cost is smaller, therefore the best performance is achieved when $T=3$.
This confirms the fact that taking all user's location in a frame into consideration can make wiser service placement decision, which gains the better performance and helps to avoid frequent migration and non-migration decision. 

Fig. 14 compares the time-averaged migration cost queue between \textbf{OSP} and \textbf{PSP} with different values of control parameter $V$. Widely, as $V$ increases, the averaged queue backlog increases in an approximately linear fashion, except for several fluctuations, which matches Theorem 1 and 2. Besides the \textbf{PSP} scheme has a better performance in queue backlog with a large value of $V$. Along with Fig. 4, the performance of latency and cost obeys the $[O(1/V),O(V)]$ trade-off. 
Meanwhile Fig. 15 plots that the change curve of the migration backlog queue tends to be stable no matter what $V$ is, which implies our proposed algorithms will satisfy the long-term cost budget. Obviously, \textbf{PSP} achieves better performance than \textbf{OSP} under the same $V$.
\begin{figure}[tbp]
	\begin{minipage}[t]{0.22\textwidth}
		\centering
	\includegraphics[height=1.75in,width=1.8in]{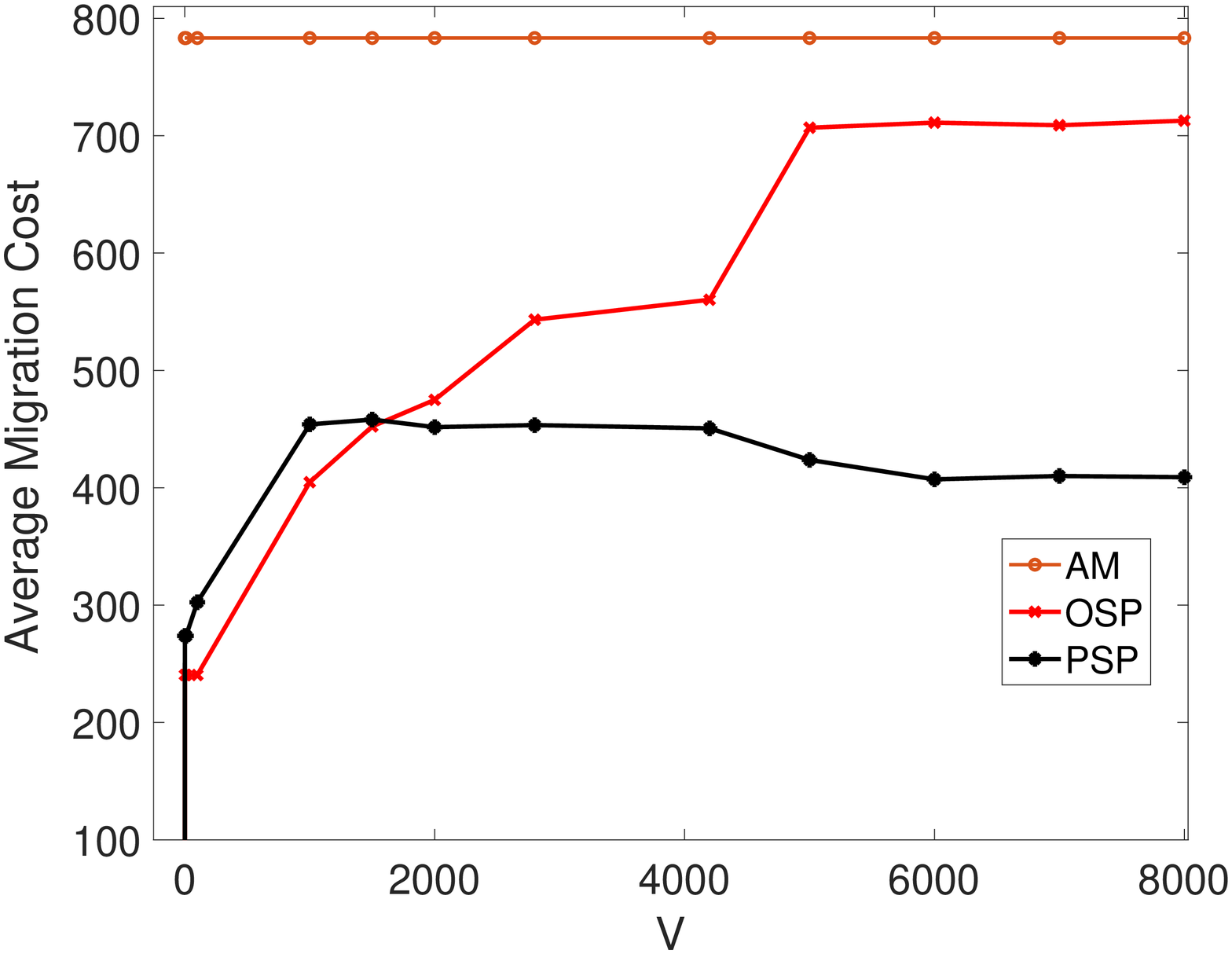}
	\caption{Average migration cost with different values of  $V$}
	\end{minipage}
\hfill
	\begin{minipage}[t]{0.23\textwidth}
		\centering
		\includegraphics[height=1.75in,width=1.8in]{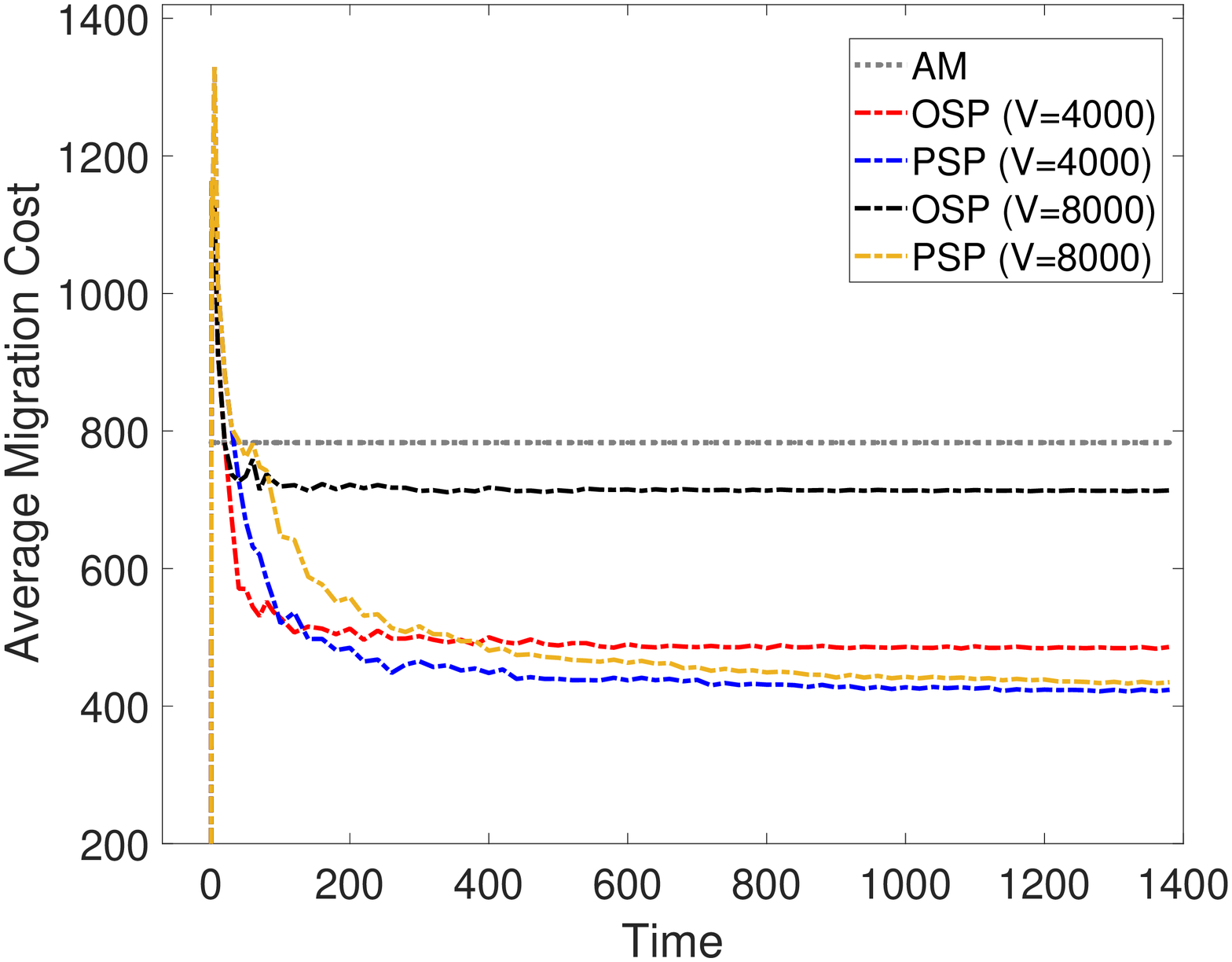}  
	\caption{Average migration cost with different values of  $V$ at different time slot}
	\end{minipage}
	\hfill
\end{figure}

\emph{Convergence of average migration cost.} Fig. 16 plots the average migration cost with different values of $V$ under our algorithms. It is worth noting that the migration cost budget is nearly half the cost of all services migration. In this situation, a large value of $V$ makes system care more about user-perceived latency, which may violate the long-term migration cost budget in finite time slots, such as $V=8000$. Obviously, the average migration cost in \textbf{PSP} is smaller and more stable than that in \textbf{OSP}. As shown in Fig. 17, with the increase of time slot, the average migration cost decreases significantly, and gradually converges to the migration cost budget when $V$ in different values.
This is because of the insufficient time slots in our simulations. As mentioned earlier, the migration cost queue  stability is equivalent to realize $\lim_{\widetilde{T}\to\infty} \mathbb E{[Q(\widetilde{T})]}/{\widetilde{T}}=0$ to ensure that the actual migration cost would beneath the budget. In Fig. 15, we know that all migration backlog queues gradually converge to some certain finite value. Thus increasing time slots can satisfy the long-term constraint. It is clearly, \textbf{PSP} achieves the better performance than \textbf{OSP} under the same $V$ and same time slots.
\section{CONCLUSION}
In this paper, we study the dynamic service performance optimization problem with long-term time-averaged migration cost budget. 
By applying Lyapunov optimization technique, we first design a one-slot reactive online service placement (\textbf{OSP})  algorithm to decompose the long-term optimization problem into a series of real-time optimization problem without requiring future information (such as user mobility). 
Aiming at a paradigm shifting from reactive to proactive by leveraging the power of prediction for performance enhancement, we further study predictive service placement with predicted near-future information. By using two-timescale Lyapunov optimization method, we propose a $T$-slot predictive service placement (\textbf{PSP}) algorithm to incorporate the prediction of user mobility. We characterize the performance bounds of \textbf{OSP} and \textbf{PSP} in terms of cost-delay trade-off theoretically. 
And we further exploit the historical queue information to add a new weight adjustment scheme for queue management named weight-update method for the \textbf{PSP} algorithm (i.e., \textbf{PSP-WU}), which greatly reduces the length of queue while improving the quality of user-perceived latency.
We conduct extensive experiments using real-world data traces, which show that our model performs effectively to reduce user-perceived latency while keeping cost consumption low and stable.

\bibliographystyle{IEEEtran}
\bibliography{ref}

\end{document}